\documentclass[graybox]{svmult}

\usepackage{type1cm}        
\usepackage{makeidx}         
\usepackage{graphicx}        
\usepackage{multicol}        
\usepackage[bottom]{footmisc}

\usepackage{newtxtext}       %
\usepackage{newtxmath}       

\usepackage{braket}
\usepackage[hyperfootnotes=false]{hyperref} 
\usepackage{cite}

\makeindex             

\newcommand{\Apeak}{A_{\mathrm{peak}}}

\begin{document}

\title*{Strong-Field Electron Dynamics in Solids}
\author{Kenichi L. Ishikawa, Yasushi Shinohara, Takeshi Sato, and Tomohito Otobe}
\institute{
Kenichi L. Ishikawa \at Department of Nuclear Engineering and Management, Graduate School of Engineering, The University of Tokyo, 7-3-1 Hongo, Bunkyo-ku, Tokyo 113-8656, Japan, \email{ishiken@n.t.u-tokyo.ac.jp}
\and
Yasushi Shinohara \at Photon Science Center, Graduate School of Engineering, The University of Tokyo, 7-3-1 Hongo, Bunkyo-ku, Tokyo 113-8656, Japan, \email{shinohara@atto.t.u-tokyo.ac.jp}
\and
Takeshi Sato \at Department of Nuclear Engineering and Management, Graduate School of Engineering, The University of Tokyo, 7-3-1 Hongo, Bunkyo-ku, Tokyo 113-8656, Japan, \email{sato@atto.t.u-tokyo.ac.jp}
\and
Tomohito Otobe \at Ultrafast Dynamics Group, National Institutes for Quantum and Radialogical Science and Technology (QST), 8-1-7, Umemidai, Kizugawa, Kyoto 619-0215, Japan, \email{otobe.tomohito@qst.go.jp}}
%
%
\maketitle

\abstract*{Solid-state materials have recently emerged as a new stage of strong-field physics and attosecond science. The mechanism of the electron dynamics driven by an ultrashort intense laser pulse is under intensive discussion. Here we theoretically discuss momentum-space strong-field electron dynamics in graphene and crystalline dielectrics and semiconductors. First, within massless Dirac fermion and tight-binding models for graphene, we rigorously derive intraband displacement and interband transition, which form the basis for understanding solid-state strong-field physics including high-harmonic generation (HHG). Then, based on the time-dependent Schr\"odinger equation for a one-dimensional model crystal, we introduce a simple, multiband, momentum-space three-step model that incorporates intraband displacement, interband tunneling, and recombination with a valence band hole. We also analyze how the model is modified by electron-hole interaction. Finally, actual three-dimensional materials are investigated. We present a time-dependent density-matrix method whose results for HHG are compared with experimental measurement results. Moreover, we describe the dynamical Franz-Keldysh effect in femtosecond time resolution, i.e., the time-dependent modulation of a dielectric function under an intense laser field, using a real-time time-dependent density functional theory.
}

\abstract{Solid-state materials have recently emerged as a new stage of strong-field physics and attosecond science. The mechanism of the electron dynamics driven by an ultrashort intense laser pulse is under intensive discussion. Here we theoretically discuss momentum-space strong-field electron dynamics in graphene and crystalline dielectrics and semiconductors. First, within massless Dirac fermion and tight-binding models for graphene, we rigorously derive intraband displacement and interband transition, which form the basis for understanding solid-state strong-field physics including high-harmonic generation (HHG). Then, based on the time-dependent Schr\"odinger equation for a one-dimensional model crystal, we introduce a simple, multiband, momentum-space three-step model that incorporates intraband displacement, interband tunneling, and recombination with a valence band hole. We also analyze how the model is modified by electron-hole interaction. Finally, actual three-dimensional materials are investigated. We present a time-dependent density-matrix method whose results for HHG are compared with experimental measurement results. Moreover, we describe the dynamical Franz-Keldysh effect in femtosecond time resolution, i.e., the time-dependent modulation of a dielectric function under an intense laser field, using a real-time time-dependent density functional theory.
}

\section{Introduction}
\label{sec:Introduction}

The Physics Nobel Prize in 2018 was awarded for groundbreaking inventions in the field of laser physics to Arthur Ashkin, G\'erard Mourou, and Donna Strickland.
Among them, the prize motivation for G.~Mourou and D.~Strickland was ``for their method of generating high-intensity, ultra-short optical pulses."
Ultrashort (typically femtosecond), intense laser pulses enabled by their invention, chirped-pulse amplification (CPA) \cite{Strickland1985OC}, have become an important tool in scientific research as well as industrial applications.
In {\it Scientific Background} \cite{SciBack2018}, the Nobel Committee for Physics at the Royal Swedish Academy of Sciences names CPA technology's major applications, among which the first is {\it strong-field physics and attosecond science} and the third {\it high-intensity lasers in industry and medicine}.

Atoms and molecules in the gas phase irradiated by a high-intensity femtosecond laser pulse exhibits highly nonlinear behavior such as above-threshold ionization, tunneling ionization, non-sequential double ionization, and high-harmonic generation (HHG) \cite{Protopapas1997RPP,Brabec2000RMP}.
The strong-field physics is a field that studies these {\it strong-field phenomena}.
HHG, in particular, represents a highly successful avenue toward an attosecond coherent light source in the extreme-ultraviolet and soft x-ray spectral ranges \cite{Popmintchev2012Nature,Chang2011,AttosecondPhysics}, which has opened new research possibilities including attosecond science \cite{Salieres2012RPP,Agostini2004RPP,Krausz2009RMP,Gallmann2013ARPC} to observe and manipulate ultrafast electron dynamics.
The gas-phase strong-field phenomena can be consistently explained by the so-called three-step model \cite{Corkum1993PRL,Kulander1993Nato}, in which an electron is first ejected by tunneling ionization by the strong field, then accelerated classically by an oscillating laser field, and finally recombines or recollides with the parent ion.

Thanks to the advent of high-intensity mid-infrared to terahertz radiation sources, solid-state materials have recently emerged as a new stage of strong-field physics and attosecond science \cite{Ghimire2019NPhys}.
In particular, many experimental observations of HHG from solids have been reported since the first discovery by Ghimire {\it et al.} \cite{Ghimire2011,Schubert_2014,Luu2015a,Vampa2015a,Hohenleutner2015a,Ndabashimiye_2016,Han2016,Garg2016,Liu2017,You2017,LangerF.2017,You2017OL,Kaneshima2018,Hirori2019}, revealing unique aspects of solid-state HHG such as linear scaling of cutoff energy with field strength \cite{Ghimire2011,Luu2015a} and multiple plateau structure \cite{Ndabashimiye_2016,You2017OL}.
In contrast to the gas-phase case, the mechanism underlying the strong-field electron dynamics in solids has turned out to be complex and depend on experimental conditions.
Among factors specific to solids are, 
\begin{itemize}
	\item Co-presence of intraband and interband transitions
	\item electron-hole interaction (e.g., exciton formation) and electron correlation (e.g., carrier scattering and excitonic molecule formation)
	\item dependence on crystal orientation and laser polarization due to crystal anisotropy 
\end{itemize}
Such complexity and diversity will make strong-field electron dynamics in solids offer even richer information on band structure and ultrafast dynamic electron correlation.
Strong-field physics and attosecond science have enabled detailed analysis on intense laser interaction with matter.
The extension of its frontier from atomic and molecular systems to solid-state materials will further advance industrial and medical applications of high-intensity lasers such as laser material processing.

In this Chapter, we present theories on momentum-space electron dynamics in graphene and crystalline dielectrics and semiconductors subject to intense laser fields. First, we discuss graphene within massless Dirac fermion and tight-binding models, where we rigorously derive intraband and interband transitions, forming the basis for understanding solid-state strong-field physics (Sect.~\ref{sec:Graphene}). 
Then, based on the time-dependent Schr\"odinger equation for a one-dimensional model crystal within a single-electron approximation, we introduce a simple, {\it multiband}, momentum-space three-step model that incorporates intraband displacement, interband tunneling, and recombination with a valence band hole. 
We further analyze electron-hole interaction effects, using the time-dependent Hartree-Fock calculations (Sect.~\ref{sec:Solid-State Three-Step Model}). 
Finally, actual three-dimensional materials are investigated. We present a time-dependent density-matrix method useful to quantitatively understand and explain experimental results (Sect.~\ref{sec:TDDM}). 
Moreover, we describe the dynamical Franz-Keldysh effect in femtosecond time resolution, i.e., the time-dependent modulation of a dielectric function under an intense laser field, using a real-time time-dependent density functional theory (Sect.~\ref{sec:DFKE}).
Summary is given in Sect.~\ref{sec:Summary}.
Hartree atomic units are used throughout unless otherwise stated.

\section{Graphene}
\label{sec:Graphene}

It is instructive to examine the laser-driven coherent electron dynamics in graphene \cite{Novoselov2004Science,Berger2004JPCB,Geim2007NM,Geim2009Science} , for which the intraband and interband transitions, key to understand solid-state HHG, can be rigorously and simply derived.
Let us consider a single-electron response in the mono-layer graphene placed in the $xy$ plane subject to normal incidence of a laser pulse with its electric field ${\bf E}(t)$ and vector potential ${\bf A}(t)=-\int {\bf E(t)} dt$ being in the graphene plane. 

\subsection{Graphene Bloch Equations (GBEs)}

The two-component wave function $\psi (t)$ of the electron with an initial wave vector of ${\bf k}$ and canonical momentum ${\bf p}=\hbar {\bf k}$ is governed by the following time-dependent Schr\"odinger equation (TDSE):
\begin{equation}
\label{eq:TDSE}
i\hbar\frac{\partial}{\partial t}\psi (t)=H(t) \psi (t),
\end{equation}
with the time-dependent Hamiltonian $H(t)$,
\begin{equation}
\label{eq:hamiltonian}
H(t)=\left(\begin{array}{cc}0 & h(t) \\h(t)^* & 0\end{array}\right)
\qquad 
h(t)=\epsilon (t) e^{-i\theta (t)},
\end{equation}
where $\epsilon (t) =|h(t)|$ and $\theta (t)=-\arg h(t)$.

\begin{figure}[tb] 
\centerline{\includegraphics[width=\textwidth]{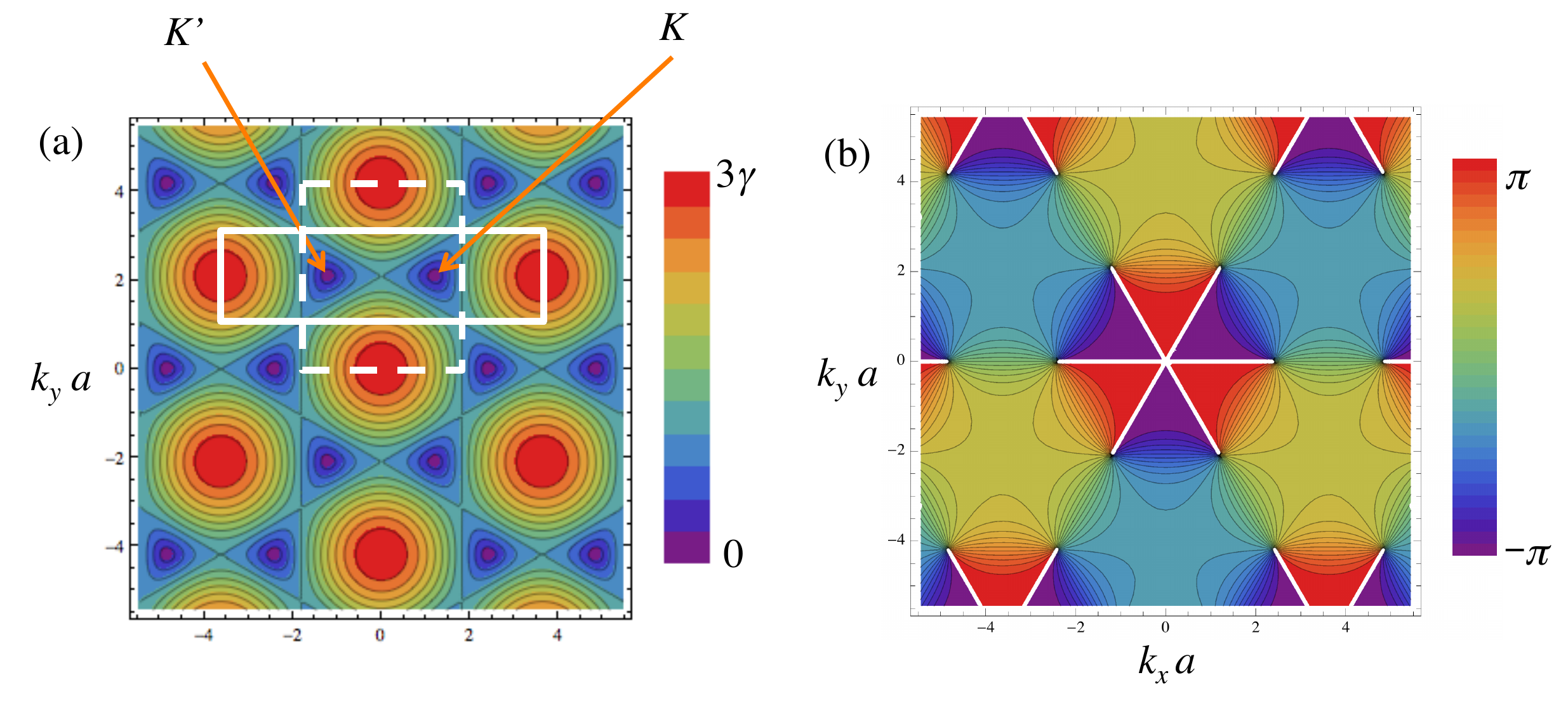}} 
\caption{(color online) Contour and false color plots of (a) $\epsilon_{\bf k}$ and (b) the principal value $-{\rm Arg}\, h_{\bf k}$, defined in the range $[-\pi,\pi)$ \cite{graphene2013NJP}. The thick solid and dashed white squares in panel (a) outline examples of simulational Brillouin zone, i.e., the area of integration in Eq.\ (\ref{eq:total_current}). The Dirac points $K$ and $K^\prime$ are located at $\left(\pm\frac{2\pi}{3\sqrt{3}}, \frac{2\pi}{3}\right)=(\pm1.2092,2.0944)$. In panel (b), the value jumps between $\pi$ and $-\pi$ on solid white lines. Reprinted from Ref.~\cite{graphene2013NJP} with permission under the Creative Commons Attribution 3.0 license.}
\label{fig:tight_binding}
\end{figure}

Within the framework of the tight-binding (TB) model of nearest-neighbor interactions \cite{CastroNeto2009RMP,Gusynin2007IJMP},
\begin{equation}
\label{eq:tbh}
h(t) = -\gamma\sum_{\alpha=1}^{3} e^{i{\boldsymbol\kappa}\cdot {\boldsymbol \delta}_\alpha},
\end{equation}
where $\gamma \approx 2.5-2.8\,{\rm eV}$ denotes the hopping energy, and ${\boldsymbol \delta}_1=a(0,1)$ and ${\boldsymbol \delta}_{2,3}=\frac{a}{2}(\pm\sqrt{3},-1)$ the locations of nearest neighbors separated by distance $a\approx 1.42\,{\rm \AA}$. 
${\boldsymbol \kappa} =  {\boldsymbol \pi}/\hbar$ is the wave vector corresponding to the kinetic momentum ${\boldsymbol \pi}(t)={\bf p}+e{\bf A}(t)$ with $e(>0)$ being the elementary charge.
${\boldsymbol \kappa}$ and ${\boldsymbol \pi}$ vary with time and describe the laser-driven intraband displacement. 
$\epsilon$ denotes the magnitude of the energy eigenvalue in the absence of the field whose value $\epsilon_{\bf k}$ for ${\bf k}$ is given by [Fig.\ \ref{fig:tight_binding} (a)],
\begin{equation}
\label{eq:epsilon_tb}
\epsilon_{\bf k}=\gamma\sqrt{3+2\cos \sqrt{3}k_x a+4\cos\frac{\sqrt{3}k_x a}{2}\cos\frac{3k_y a}{2}}.
\end{equation}

\begin{figure}[tb]
\centerline{\includegraphics[width=6cm]{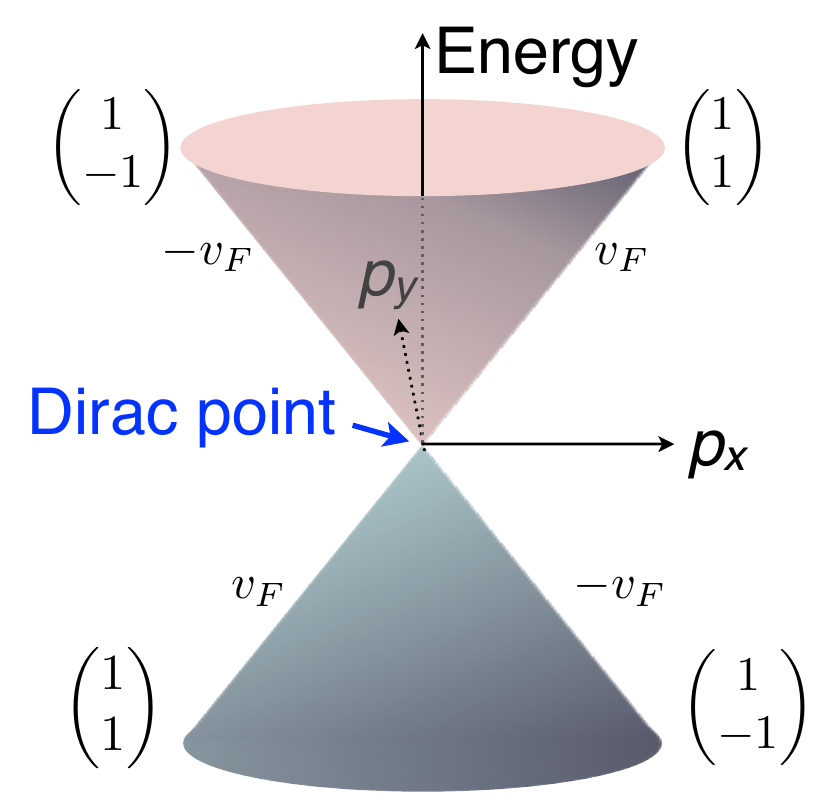}} 
\caption{Electronic band structure of monolayer graphene near the Dirac point (Dirac cone) within the massless Dirac fermion picture. Reprinted with permission from Ref.~\cite{graphene2010PRB}. Copyright 2010 by American Physical Society.}
\label{fig:dirac-cone}
\end{figure}

If we resort to the massless Dirac fermion (MDF) picture \cite{CastroNeto2009RMP} applicable near the Dirac point and take the Dirac point as the origin of ${\boldsymbol \pi}$ (Fig.~\ref{fig:dirac-cone}) , $h(t)$ and $\epsilon (t)$ are simplified to, 
\begin{equation}
\label{eq:h-MDF}
h(t) = v_F(\pi_x-i\pi_y) = v_F[(p_x+eA_x)-i(p_y+eA_y)],
\end{equation}
and
\begin{equation}
\label{eq:eigenvalueMDFnormal}
\epsilon (t) = v_F |\boldsymbol\pi (t)| = v_F\hbar |\boldsymbol\kappa (t)|,
\end{equation}
respectively, with $v_F=\frac{3\gamma a}{2\hbar}\approx c/300$.
It should be noted that TDSE Eq.~(\ref{eq:TDSE}) has a form similar to the Dirac equation but is different from the original Dirac equation in that the fermion mass is zero, which leads to the two-component, instead of four-component, wave function.
We also notice that $\theta$ becomes the directional angle of $\boldsymbol\kappa$: $\kappa_x=|\boldsymbol\kappa| \cos \theta$ and $\kappa_y=|\boldsymbol\kappa| \sin \theta$ (around ${\bf K}$), $-|\boldsymbol\kappa| \sin \theta$ (around ${\bf K}^\prime$).

In the field-free case, whether we may use the TB or MDF pictures, the TDSE has the following two solutions:
\begin{equation}
\psi(t)=\frac{1}{\sqrt{2}}\exp \left(\mp i\frac{\epsilon}{\hbar}t\right)
\left(\begin{array}{c}e^{-\frac{i}{2}\theta} \\\pm e^{\frac{i}{2}\theta}\end{array}\right)
\end{equation}
with their energy eigen values are $\pm\epsilon$.
The upper sign refers to the upper band (electron band), and the lower sign to the lower band (hole band).

Let us now turn on the laser pulse and express the wave function as a superposition,
\begin{equation}
\label{eq:ansatz1}
\psi(t)=c_+(t)\psi_+(t)+c_-(t)\psi_-(t),
\end{equation}
of the instantaneous upper and lower band states (Volkov states),
\begin{equation}
\label{eq:ansatz2}
\psi_\pm(t)=\frac{1}{\sqrt{2}}\exp \left[\mp i\Omega(t)\right]
\left(\begin{array}{c}e^{-\frac{i}{2}\theta(t)} \\ \pm e^{\frac{i}{2}\theta(t)}\end{array}\right)
\end{equation}
with the instantaneous temporal phase or dynamical phase $\Omega (t)$ defined as,
\begin{equation}
\label{eq:phase}
\Omega(t) = \int \frac{\epsilon (t)}{\hbar}\, dt.
\end{equation}
We find that Eq.\ (\ref{eq:ansatz1}) is indeed the exact solution of the TDSE Eq.~(\ref{eq:TDSE}) if the expansion coefficients $c_{\pm}(t)$ satify the equations of motion:, 
\begin{equation}
\label{eq:variation}
\dot{c}_\pm(t)=\frac{i}{2}\dot{\theta}(t)c_\mp(t)e^{\pm 2i\Omega(t)}.
\end{equation}
Introducing the population difference $n=|c_+|^2-|c_-|^2$ between the two band and the interband coherence $\rho=c_+c_-^*$, we can transform Eq.\ (\ref{eq:variation}) into the graphene Bloch equations \cite{graphene2010PRB}:
\begin{align}
\label{eq:GBE2}
\dot{n}&=-i\,\dot{\theta}(t)\rho (t) \, e^{-2i\Omega(t)}+{\rm c.c.},\\
\label{eq:GBE1}
\dot{\rho}&=-\frac{i}{2}\dot{\theta}(t) n(t) e^{2i\Omega(t)}.
\end{align}
To take account of the Fermi distribution at finite temperature $T$, we solve Eqs.\ (\ref{eq:GBE2}) and (\ref{eq:GBE1}) under initial conditions $n=F(p)-F(-p)$ and $\rho=0$, where $F(p)=\{1+\exp [(\epsilon (p)-\mu)/k_BT]\}^{-1}$ is the Fermi-Dirac function, where $\mu$ and $k_B$ denote the chemical potential and Boltzmann constant, respectively. 

Through $\Omega (t)$ and $\theta (t)$, Eq.~(\ref{eq:ansatz2}) incorporates the field-induced intraband dynamics (transition) of the electron that changes its kinetic momentum ${\bf p}+e{\bf A}(t)$ following the acceleration theorem.
On the other hand, Eqs.\ (\ref{eq:variation})--(\ref{eq:GBE1}) indicate that the electron undergoes interband transitions while retaining coherence.
Thus, the GBEs are physically more transparent than the TDSE.

If we defined $\theta(t)$ by the principal value $\theta(t)=-{\rm Arg}\,h(t)$ with ${\rm Arg}\,z\in (-\pi,\pi]$, as plotted in Fig.\ \ref{fig:tight_binding} (b), $\theta(t)$ would undergo $2\pi$ jumps on the white lines linking Dirac points. 
When we follow the electron dynamics based on the GBEs, instead, we need to define $\theta(t)=-\arg h(t)$, with $\arg z = {\rm Arg} \,z + 2\pi n$ ($n$ is an integer), in such a way that it varies continuously along the path of ${\boldsymbol\kappa} (t)$.
It should be noticed that if ${\boldsymbol\kappa} (t)$ takes a ${\bf k}$-space trajectory surrounding a Dirac point, $\psi_\pm (t)$ acquires a geometrical phase of $\pi$, in addition to the dynamical phase $\Omega(t)$. 
Thus, Berry's phase \cite{Berry1984,Zhang2005Nature} is incorporated in the GBEs.

\begin{figure}[tb]
\centerline{\includegraphics[width=8.3cm]{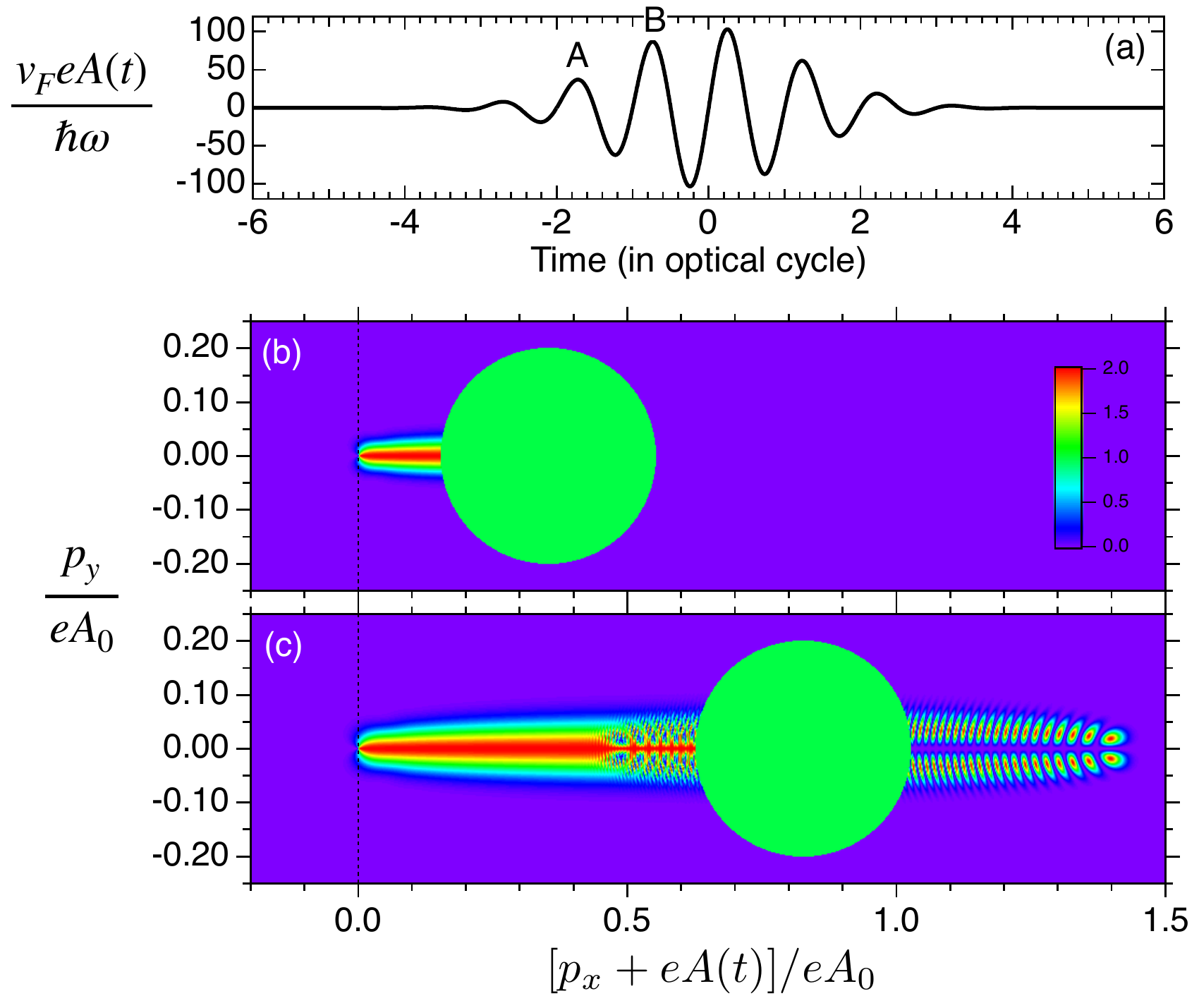}} 
\caption{(a) Normalized vector potential $v_FeA(t)/\hbar\omega$ of the incident optical pulse. (b) and (c) Carrier occupation distribution calculated for the moment labeled as {\it A} and {\it B}, respectively, in panel (a). Reprinted with permission from Ref.~\cite{graphene2010PRB}. Copyright 2010 by American Physical Society.}
\label{fig:carrier_occupation}
\end{figure}

It should be noticed that the whole dynamics within the MDF model is invariant under multiplication of quantities of energy dimension, $\hbar\omega$, $v_Fp$, $v_FeA$, $\mu$, $k_BT$, and $\hbar/t$, by a common factor. In the weak-field limit, one can derive the universal conductivity $e^2/4\hbar$ from Eqs.\ (\ref{eq:jcx}) -- (\ref{eq:total_current_MDF}) \cite{graphene2013NJP}.

An example of carrier occupation distribution calculated within the MDF model is visualized in Fig.~\ref{fig:carrier_occupation} for the case where $T=0$, $\mu = v_F e A_0/5$, and the vector potential $A(t)$ is assumed to be a sine pulse with a Gaussian intensity envelope whose full-width-at-half-maximum width corresponds to two optical cycles, and peak amplitude $A_0$ satisfies $\hbar\omega/v_FeA_0=9.46\times 10^{-3}$ [Fig.~\ref{fig:carrier_occupation} (a)].
Figure \ref{fig:carrier_occupation} (b) and (c) show carrier occupation distribution at moments {\it A} and {\it B} marked in Fig.~\ref{fig:carrier_occupation} (a).
The green circle represents the electrons originally in the upper band around the Dirac point (note that $\mu>0$), which undergoes intraband displacement.
In addition, outside the green circle, part of the electrons initially in the lower band transfer to the upper band through interband transitions.
An electron with a given initial momentum may transfer to the other band each time it passes near the Dirac point in the oscillating laser field, thus split into different quantum pathways to reach the conduction band.
Their interference is clearly seen in Fig.~\ref{fig:carrier_occupation} (c).
Such an interference effect has been experimentally observed and controlled by Higuchi {\it et al.} \cite{Higuchi2017Nature}.


\subsection{Electric Current and Harmonic Generation}

The single-electron electric current is given by $-e{\bf j}_e$, where ${\bf j}_e$ is defined as,
\begin{equation}
\label{eq:single_electron_current}
{\bf j}_e = \psi^\dagger (t)\frac{\partial H}{\partial {\boldsymbol\pi}} \psi (t).
\end{equation}
For the case of the TB model, using $\rho (t)$ and $n(t)$, we obtain the following explicit form: 
\begin{equation}
\label{eq:je-TB}
{\bf j}_e = \frac{\gamma}{\hbar}\sum_{\alpha=1}^{3}\left[n\,\sin\left({\boldsymbol\kappa}\cdot {\boldsymbol \delta}_\alpha+\theta\right) 
-i\left\{\rho\, e^{-i2\Omega}\cos\left({\boldsymbol\kappa}\cdot {\boldsymbol \delta}_\alpha+\theta\right)-{\rm c.c.}\right\}\right]{\boldsymbol \delta}_\alpha.
\end{equation}
Then, to obtain the macroscopic electric current ${\bf J}(t)$ generated by the laser field, we calculate the carrier current ${\bf j}_c$ by replacing $n$ with the carrier occupation $n+1$ in Eq.\ (\ref{eq:je-TB}) then integrate $-e{\bf j}_c$ over the honey-comb lattice Brillouin zone [Fig.\ \ref{fig:tight_binding} (a)] as, 
\begin{equation}
\label{eq:total_current}
{\bf J}(t) = -\frac{g_s e}{(2\pi)^2}\int_{\rm BZ}{\bf j}_c(t)d{\bf k},
\end{equation}
where $g_s=2$ denotes the spin-degeneracy factor. 

The expressions in the MDF picture are simpler. Each component of the carrier current is written as,
\begin{align}
\label{eq:jcx}
j_{c,x}&=v_F\left[ (n+1)\cos\theta+i\sin\theta\{\rho e^{-2i\Omega}-{\rm c.c.}\}\right],\\
\label{eq:jcy}
j_{c,y}&=v_F\left[ (n+1)\sin\theta-i\cos\theta\{\rho e^{-2i\Omega}-{\rm c.c.}\}\right].
\end{align}
Then, the macroscopic electric current ${\bf J}(t)$ is given by, 
\begin{equation}
\label{eq:total_current_MDF}
{\bf J}(t) = -\frac{g_s g_v e}{(2\pi\hbar)^2} \int {\bf j_c}(t)d{\bf p} = -\frac{g_s g_v e}{(2\pi)^2} \int {\bf j_c}(t)d{\bf k},
\end{equation}
with $g_v=2$ being the valley-degeneracy factor.

One can calculate the intensity spectrum $I(\omega)$ of harmonic generation using the Fourier transform $\hat{\bf J}(\omega)$ of ${\bf J}(t)$ by,
\begin{equation}
\label{eq:harmonics}
I(\omega)\propto |\omega\hat{\bf J}(\omega)|^2.
\end{equation}
In Eqs.~(\ref{eq:jcx}) and (\ref{eq:jcy}) we can identify the contribution to harmonic generation from the oscillating interband polarization $\rho (t) e^{-2i\Omega (t)}$ and the intraband transition $\theta (t)$ as well as the temporal variation in population $n(t)$.
Among them, the population variation contributes only to the first term, and the interband polarization only to the second, whereas both terms contain the contribution from the intraband transition.

\begin{figure}[tb]
\centerline{\includegraphics[width=8.3cm]{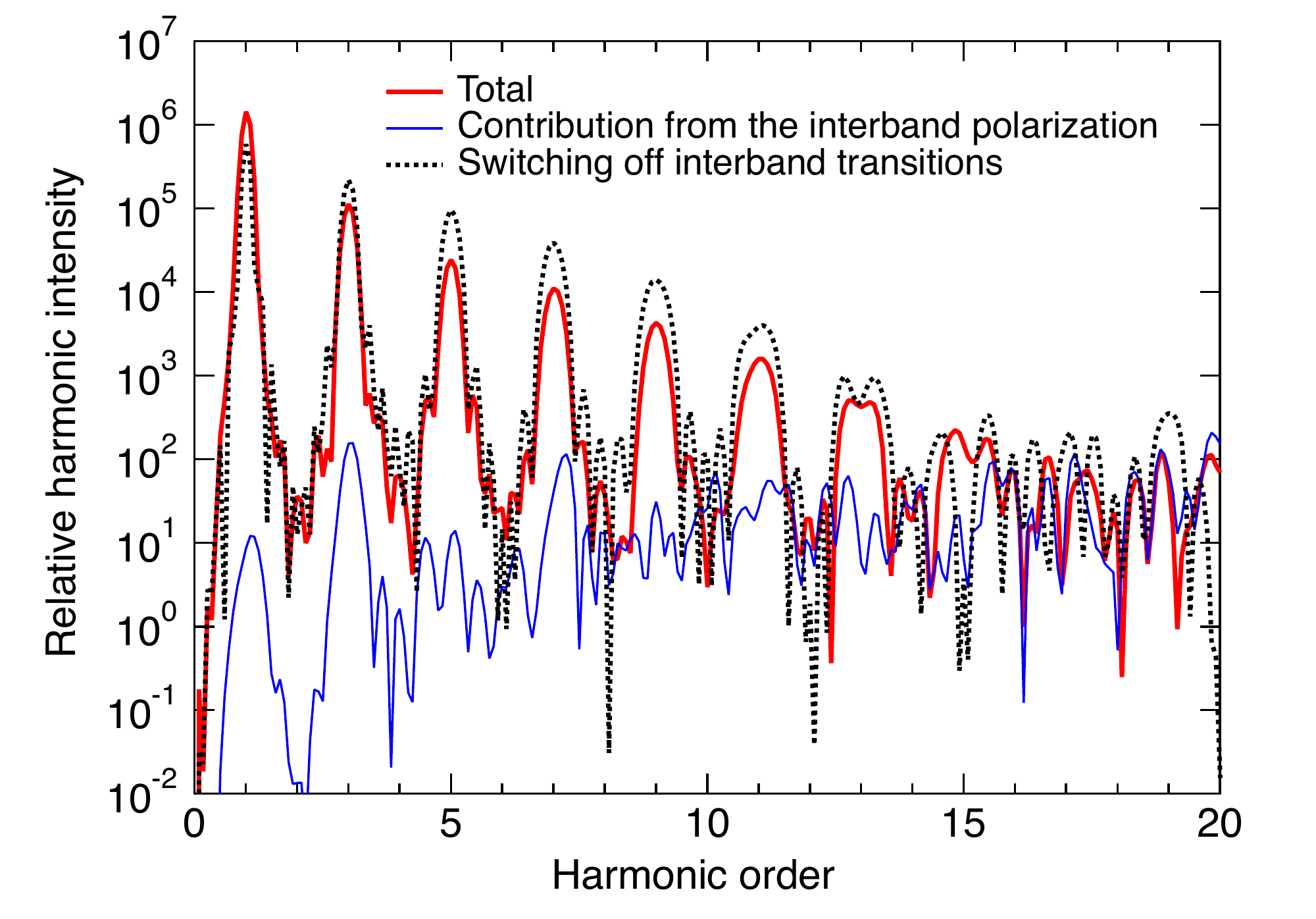}} 
\caption{Harmonic intensity spectra for the case of Fig.\ \ref{fig:carrier_occupation}. Thick solid line: total spectrum calculated with Eq.\ (\ref{eq:harmonics}), thin solid line: contribution from the interband polarization, thick dotted line: calculated by switching off the interband transitions. Reprinted with permission from Ref.~\cite{graphene2010PRB}. Copyright 2010 by American Physical Society.}
\label{fig:harmonics}
\end{figure}

In Fig.\ \ref{fig:harmonics} we plot the harmonic spectrum for the case of Fig.\ \ref{fig:carrier_occupation}.
Curiously, the harmonic intensity (red thick solid line) is reduced compared with the case of the pure intraband dynamics (thick dotted line) \cite{graphene2010PRB}, obtained by switching off interband transitions in the calculation.
Thus, in spite of the small contribution from the interband polarization itself (thin solid lines in Fig.\ \ref{fig:harmonics}), i.e., the second term of Eq.\ (\ref{eq:jcx}), the interband dynamics strongly modify the optical response of graphene, relaxing nonlinearity.
Nevertheless, harmonic generation of up to the thirteenth order can be seen, revealing high nonlineality; high-harmonic generation from graphene has been experimentally observed \cite{Bowlan2014PRB,Hafez2018Nature}.

\section{Solid-State Three-Step Model}
\label{sec:Solid-State Three-Step Model}

What characterizes high-harmonic generation is that its spectrum consists of a {\it plateau} where the harmonic intensity is nearly constant over many orders and a sharp {\it cutoff}.
The gas-phase HHG can be intuitively and even quantitatively captured by the so-called three-step model \cite{Corkum1993PRL,Kulander1993Nato}, in which an electron is first ejected by tunneling ionization by the strong field, then accelerated classically by an oscillating laser field, and finally radiatively recombines with the parent ion emitting a harmonic photon.
The cutoff energy, i.e., the maximal harmonic photon energy $E_c$ is given by,
\begin{equation}
\label{eq:gas-phase-cutoff}
E_c = I_p + 3.17 U_p,
\end{equation}
where $I_p$ is the ionization potential of the target atom or molecule, and $U_p  {\rm [eV]}= E_0^2/4\omega^2=9.337 \times 10^{-14}~I~{\rm [W/cm^2]} ~(\lambda~[\mu {\rm m}])^2$ the ponderomotive energy, with $E_0$, $I$, $\omega$, and $\lambda$ being the strength, intensity, angular frequency, and wavelength of the driving field, respectively.
Hence, the cutoff energy is roughly proportional to the square of the laser electric field strength.

High-harmonic generation from solid-state materials is, on the other hand, quite distinct from its gas-phase counterpart, exhibiting unique aspects such as linear scaling of cutoff energy with field strength \cite{Ghimire2011,Luu2015a} and multiple plateau formation \cite{Ndabashimiye_2016,You2017OL}, to name only a few.   
The comprehensive mechanism underlying solid-state HHG is under active investigation; both real-space, as in the gas-phase case, and momentum-space pictures are on the market.
In this Section, using a one-dimensional (1D) model periodic crystal, we present a solid-state momentum-space three-step model \cite{Ikemachi2017PRA,Wu2016PRA,Du2017OE,Ikemachi2018PRA} that considers electron dynamics across multiple bands, incorporating field-induced intraband displacement, interband tunneling, and recombination with the valence-band (VB) hole, suitable to discuss harmonic generation from interband polarization.
We first describe an independent-electron picture (Subsec.~\ref{subsec:Independent Electron Approximation}) \cite{Ikemachi2017PRA} and then discuss electron-hole interaction effects \cite{Ikemachi2018PRA}.

\subsection{Independent-Electron Approximation}
\label{subsec:Independent Electron Approximation}

\begin{figure}[tb]
\centering
\includegraphics[width=0.6\linewidth]{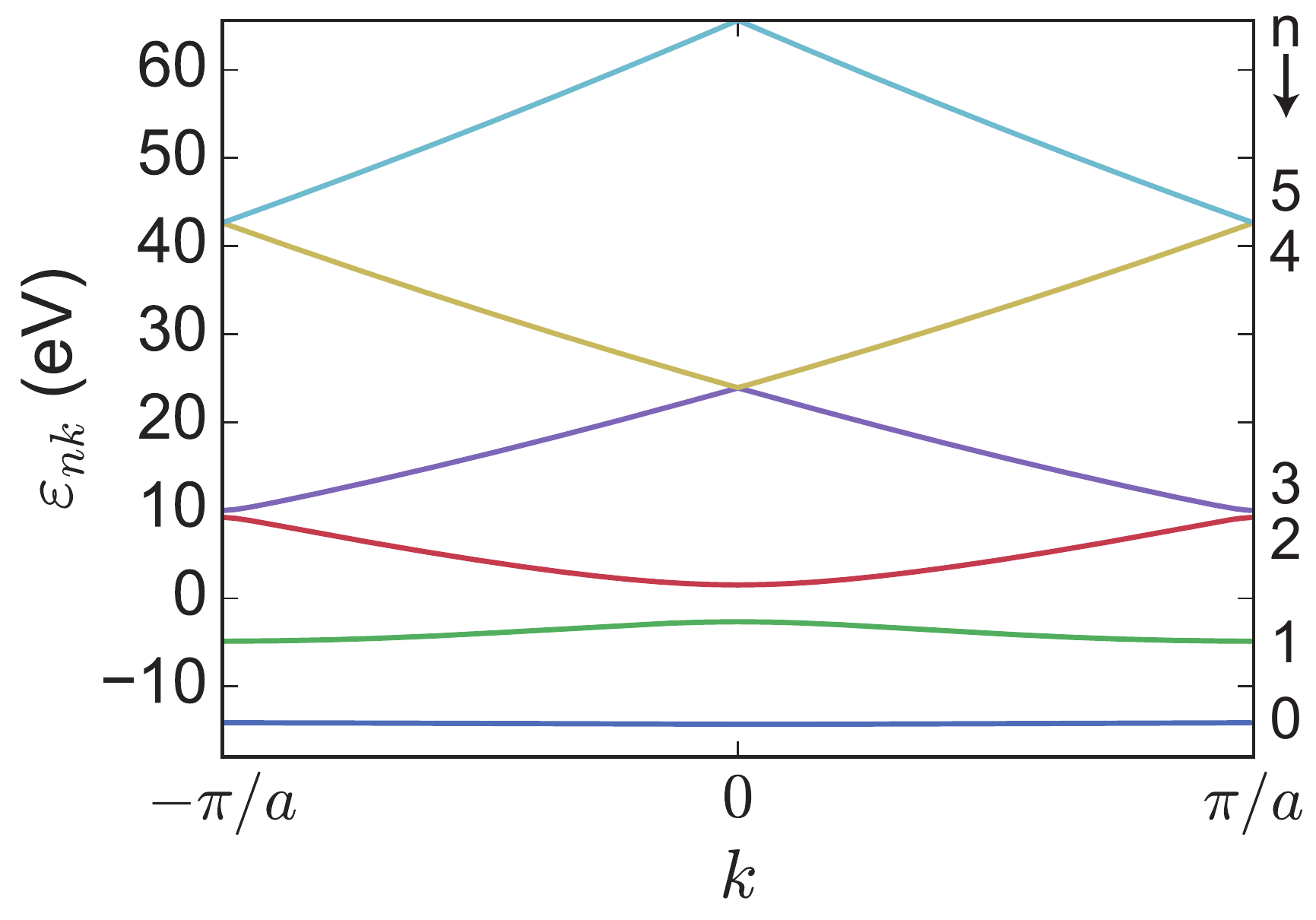}
\caption{
Two valence bands ($n=0, 1$) and first four conduction bands ($n=2, \dots, 5$) of the field-free Hamiltonian.
 The integers on the right axis are the band indices $n$. Reprinted with permission from Ref.~\cite{Ikemachi2017PRA}. Copyright 2017 by American Physical Society.
}
\label{fig:band}
\end{figure}

We consider the electron dynamics in a 1D model crystal along linear laser polarization, assuming that VBs are initially fully occupied across the whole Brillouin zone (BZ), as is usually the case for wide-band-gap semiconductors.
Let us calculate harmonic spectra, based on the effective TDSE for each independent electron,
\begin{equation}
i \frac{\partial}{\partial t} \psi_{nk}(x, t) = \hat{H}(t) \psi_{nk}(x, t)
= \left\{\frac{1}{2}\left[\hat{p} + A(t)\right]^2 + V(x) \right\} \psi_{nk}(x, t),
\label{eq:TDSE in general}
\end{equation}
within the velocity gauge for the electron initially located in band $n$ with a crystal momentum $k$, where $\hat{p}=-i\frac{\partial}{\partial x}$, $V(x)$ denotes the periodic single-electron effective potential with lattice constant $a$, i.e., $V(x+a) = V(x)$.
As the initial state of the time-dependent wave function $\psi_{nk}(x, t)$, we take the Bloch function $\phi_{nk}$, i.e., the eigenstate of
the field-free Hamiltonian
with the energy eigen value $\varepsilon_{nk}$.
We use the Mathieu-type potential \cite{Wu2015} given by,
\begin{equation}
V(x) = -V_0 \left[ 1 + \cos(2 \pi x / a)\right], 
\end{equation}
with $V_0 = 0.37$ and $a = 8$, which supports a band structure (Fig. \ref{fig:band}) with minimum band gap 4.2 eV at $k=0$, while the first and second CBs approach each other at the Bragg plane ($k=\pm \frac{\pi}{a}$).
We assume that the two VBs ($n=0, 1$ in Fig. \ref{fig:band}) are initially filled across the whole BZ.

Rather than expand the wave functions with basis functions, we resort to direct numerical integration of the TDSE Eq.~(\ref{eq:TDSE in general}) in real space.
One of the advantages of the velocity gauge is that the Hamiltonian retains lattice periodicity even under the action of the laser pulse. 
As a consequence, the initial crystal momentum $k$ is always a good quantum number, and, thus, we can solve the TDSE for each $k$ independently.
Using Bloch's theorem, the wave function $\psi_{nk}(x, t)$ can be factorized as,
\begin{equation}
\psi_{nk}(x,t) = e^{ikx}u_{nk}(x,t). \label{eq:Bloch's theorem} 
\end{equation}
where $u_{nk}(x,t)$ satisfies $u_{nk}(x+a,t) = u_{nk}(x, t)$.
The substitution of Eq.~(\ref{eq:Bloch's theorem}) into Eq.~(\ref{eq:TDSE in general}) leads the equation of motion for $u_{nk}(x, t)$,
\begin{equation}
 i \frac{\partial}{\partial t}u_{nk}(x, t) = 
  \left\{ \frac{1}{2} \left[\hat{p} + k + A(t) \right]^2 + V(x) \right\} u_{nk}(x, t). \label{eq:Schrodinger equation for u(x)}
\end{equation}
This is to be solved only within the unit cell $x\in [0, a]$, which brings substantial computational-cost reduction.
It is interesting to notice that $k + A(t)$ in Eq.~(\ref{eq:Schrodinger equation for u(x)}) automatically accounts for the intraband dynamics \cite{Kittel1987,Krieger1986} and that Eq.~(\ref{eq:Schrodinger equation for u(x)}) couples different bands.
For a given pair of $(n, k)$, we numerically integrate the equation of motion (\ref{eq:Schrodinger equation for u(x)}),
using the finite difference method with the grid spacing $0.53$ a.u., time step size $2.67\times10^{-4}$ fs $= 1.10\times10^{-2}$ a.u., and the number $N$ of $k$-points 141.

\begin{figure}[tb]
 \centering
 \includegraphics[width=0.7\linewidth]{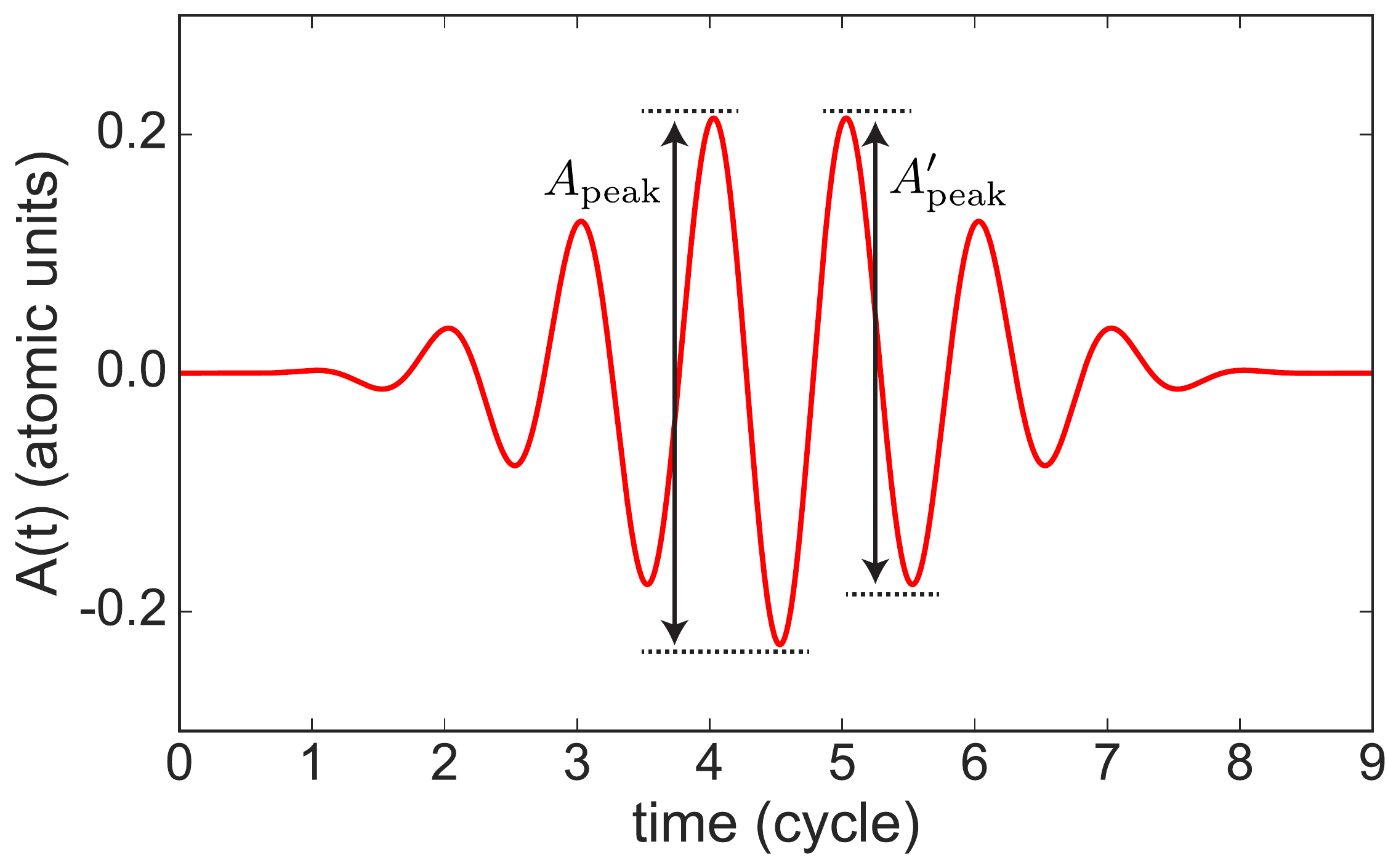}
 \caption{The waveform of the vector potential $A(t)$ of the laser pulse with  $E_0 = 1.65$ V/nm and $\tau = 99.66$ fs. The maximum and the second maximum peak-to-valley amplitude $\Apeak$ and $\Apeak^{\prime}$ are defined as depicted in the figure. Reprinted with permission from Ref.~\cite{Ikemachi2017PRA}. Copyright 2017 by American Physical Society. \label{fig:field and current}}
\end{figure}

We calculate the expectation value of velocity to obtain the contribution to the field-induced current from each $(n,k)$,
\begin{equation}
j_{nk}(t) = \bra{\psi_{nk}(t)} \hat{p} + A(t) \ket{\psi_{nk}(t)}
 = \int_{0}^{a} u_{nk}^{\ast}(x,t) \left[ \hat{p} + k + A(t)\right] u_{nk}(x,t) dx.
\label{eq:each current}
\end{equation}
Then, we obtain the total current by summing $j_{nk}$ over the initial band indices $n(=1, 2)$ and initial crystal momenta $k$,
\begin{equation}
j(t) = \frac{1}{Na} \sum_{nk} j_{nk}(t).
\end{equation}
It should be remembered that $n$ and $k$ denote the band index and crystal momentum, respectively, that the electron initially occupies.
The harmonic spectrum is calculated as the modulus square of the Fourier transform of $j(t)$.
Before applying the Fourier transform, we multiply $j(t)$ by a mask function $W(t) = \sin^4(t/\tau)$ of the same form as the field envelop in order to suppress the current remaining after the pulse.

\begin{figure}[tb]
\includegraphics[width=0.7\linewidth]{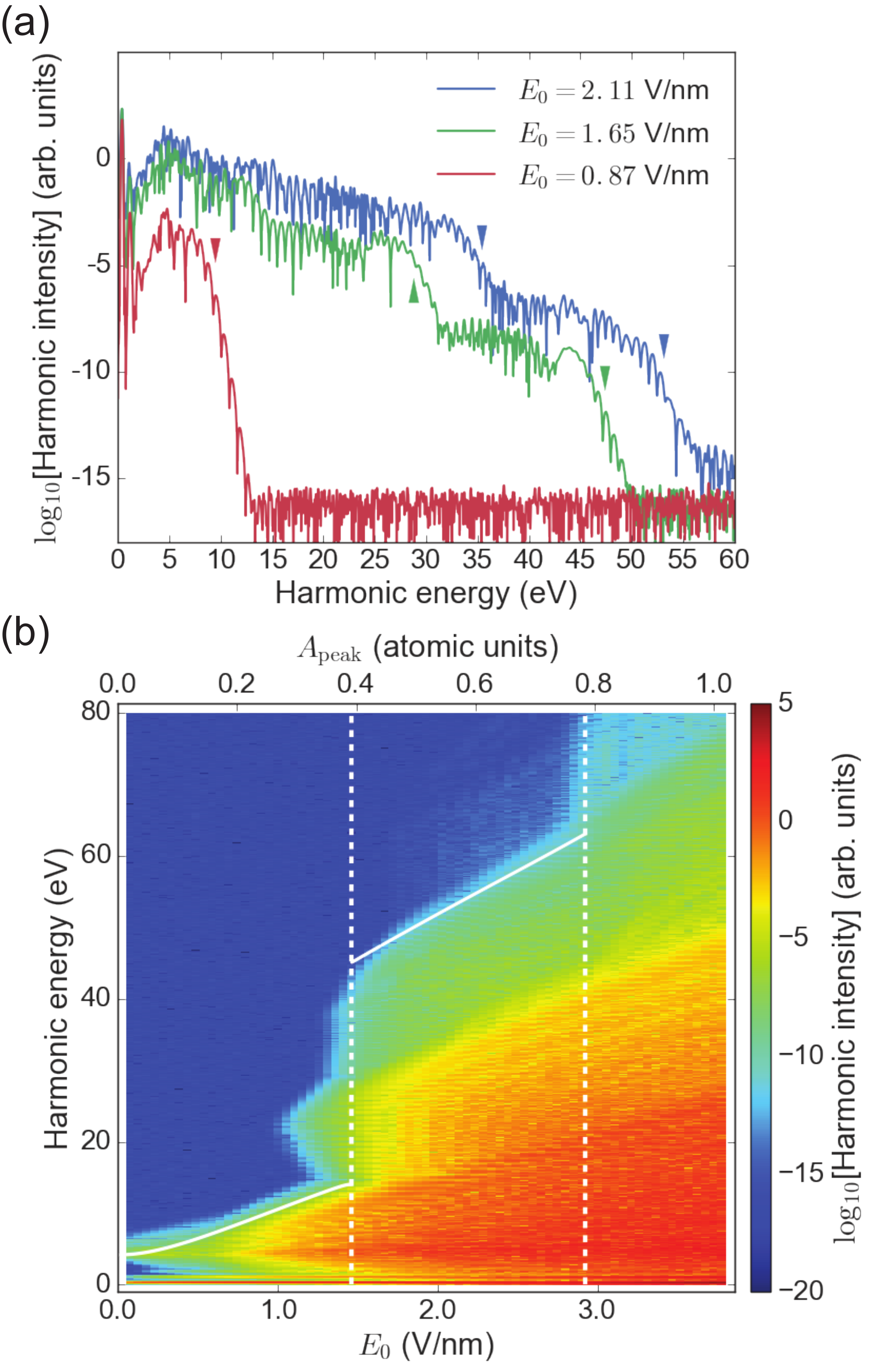}
\centering
\caption{
 (a) High harmonic spectra for $E_0 = 0.87$ V/nm (red (lower) line), $E_0 = 1.65$ V/nm (green (middle) line), and  $E_0 = 2.11$ V/nm (blue (upper) line).
 Arrowheads indicate the positions given by $\Delta\varepsilon_{21}(\Apeak)$ for $E_0 = 0.87$ V/nm (red), and Eqs.~(\ref{144641_16Nov16}) and $\Delta\varepsilon_{41}\left(\frac{\pi}{a}-\Apeak\right)$ for $E_0 = 1.65$ (green) and $2.11$ (blue) V/nm.
 (b) False-color representation of the harmonic spectra as functions of $E_0$.
 $\Apeak$ corresponding to $E_0$ is shown on the top axis in the atomic unit.
 The two vertical white dashed lines represent $\Apeak = \pi / a$ and $\frac{2\pi}{a}$.
 The two white solid lines represent the cutoff energy positions given by $\Delta\varepsilon_{21}(\Apeak)$ for $0 < \Apeak < \frac{\pi}{a}$, and Eq.~(\ref{144641_16Nov16}) for $\frac{\pi}{a} < \Apeak < \frac{2\pi}{a}$. Reprinted with permission from Ref.~\cite{Ikemachi2017PRA}. Copyright 2017 by American Physical Society.
}
\label{fig:harmonic spectrum}
\end{figure}

We specifically consider a laser electric field of a form $E(t) = E_0\sin^4( t  / \tau) \sin[\omega (t - \pi\tau/2)]$ for $t \in [0, \pi \tau]$ and $E(t) = 0$ for $t \not\in[0, \pi \tau]$, where $E_0, \tau$ denote the peak electric field amplitude and a measure of pulse width, respectively (Fig. \ref{fig:field and current}).
Figure \ref{fig:harmonic spectrum} (a) shows the high harmonic spectra for a central wavelength 3200 nm, $\tau = 96.66$ fs corresponding to a full-width-at-half-maximum (FWHM) duration of 48 fs , and several field amplitudes.
We immediately notice that, whereas the spectrum for $E_0 = 0.87$ V/nm has a single plateau and cutoff,
those for $E_0 = 1.65$ and $2.11$ V/nm have two additional plateaus of lower intensity.
Moreover, the transition from the single- to multiple-plateau structure takes place not gradually but suddenly [Fig. \ref{fig:harmonic spectrum}(b)];
while the cutoff energy increases smoothly and quasi-linearly with $E_0$ up to $\approx 1.4$ V/nm, second and third plateaus suddenly appear there, and the cutoff jumps up from 15 eV to 45 eV.
This result qualitatively reproduces the previously reported unique features of solid-state HHG \cite{McDonald2015,Wu2015,Ndabashimiye_2016}.
Also, another cutoff jump is seen at $E_0 \approx$ 2.8 V/nm.

We define the maximum peak-to-valley amplitude $\Apeak$ of $A(t)$ (see Fig.~\ref{fig:field and current}) and put it on the top axis of Fig.~\ref{fig:harmonic spectrum}(b).
Then, we find that the jump-up positions interestingly satisfy the condition that $\Apeak = \frac{\pi}{a} = 0.393$ a.u. and $\frac{2\pi}{a} = 0.786$ a.u. [vertical white dashed lines in Fig. \ref{fig:harmonic spectrum}(b)].
Note that $\Apeak$ characterizes the largest crystal momentum gain in the intraband dynamics and that $\frac{\pi}{a}$ is the distance from the $\Gamma$ point to the first-BZ edge (Fig.~\ref{fig:band}).

The following simple model \cite{Vampa2014,Vampa2015,Ikemachi2017PRA,Du2017OE} explains the above findings as well as cutoff positions and the time-frequency structure of HHG:
Its essential ingredients are summarized as follows:
\begin{enumerate}
 \renewcommand{\labelenumi}{(\roman{enumi})}
 \item {\it Tunneling ionization}: each electron is tunnel ionized to an upper band predominantly at the minimum band gap to a first approximation, e.g., from band 1 to 2 at $k=0$ and from 2 to 3 at the BZ edge in the present model crystal.
 \item {\it Intraband acceleration}: the electron is displaced in the momentum space (laser-driven intraband dynamics), following the acceleration theorem \cite{Kittel1987,Krieger1986}) $k(t) = k_0 + A(t)$ with $k_0$ being the initial crystal momentum. The resulting oscillating current leads to photoemission ({\it intraband} contribution to HHG).
 \item {\it Interband recombination}: the electron emits a photon when it undergoes an interband transition to the initial band, i.e., recombination with the valence-band hole ({\it interband} contribution to HHG).
       The photon energy is given by the particle-hole energy 
 \begin{equation}
 	\Delta\varepsilon_{n(t)n_0}(k(t)) = \varepsilon_{n(t)k(t)}-\varepsilon_{n_0k(t)},
 \end{equation}
between the band $n(t)$ where the electron is located at $t$ and the initial band $n_0$.
\end{enumerate}

Comprising tunneling ionization, acceleration, and recombination, this model can be viewed as a solid-state, momentum-space counterpart of the familiar coordinate-space three-step model \cite{Corkum1993PRL,Kulander1993Nato} of gas-phase HHG.
Nevertheless, there are important differences:
\begin{itemize}
 \item All the electrons in the VB undergo the intraband acceleration (ii) together \cite{Kittel2004,Ashcroft1976} even before the first tunneling.
       Thus, VB electrons starting from not only $k_0=0$ but also any arbitrary initial momenta $k_0$ are considered \footnote{This does not violate the Pauli exclusion principle, since all the electrons in the VB move uniformly together \cite{Kittel2004,Ashcroft1976}, and thus, no $(n, k)$ point is occupied simultaneously by more than one electron at any time.}.
 \item Electrons can climb up to higher and higher bands by repeating (i) and (ii).
 \item Not only (iii) but also (ii) contribute to harmonic generation, while harmonic photons are emitted only upon recombination in the gas phase \cite{Ghimire2012,Hawkins2013,Vampa2014,Luu2015a,Wu2015}. Thus, there are intraband and interband contributions to solid-state HHG. They can be rigorously derived for the case of graphene \cite{graphene2010PRB,graphene2013NJP}, as discussed in Sec.~\ref{sec:Graphene}.
 \item (iii) can take place at any time, in principle, while, in the gas phase, the electron can recombine with the parent ion only at the position of the latter.
\end{itemize}
This electron dynamics is conceptually similar to that in graphene \cite{graphene2010PRB,graphene2013NJP,Higuchi2017Nature} (Sect.~\ref{sec:Graphene}).

\begin{figure}[tb]
 \centering
 \includegraphics[width=\linewidth]{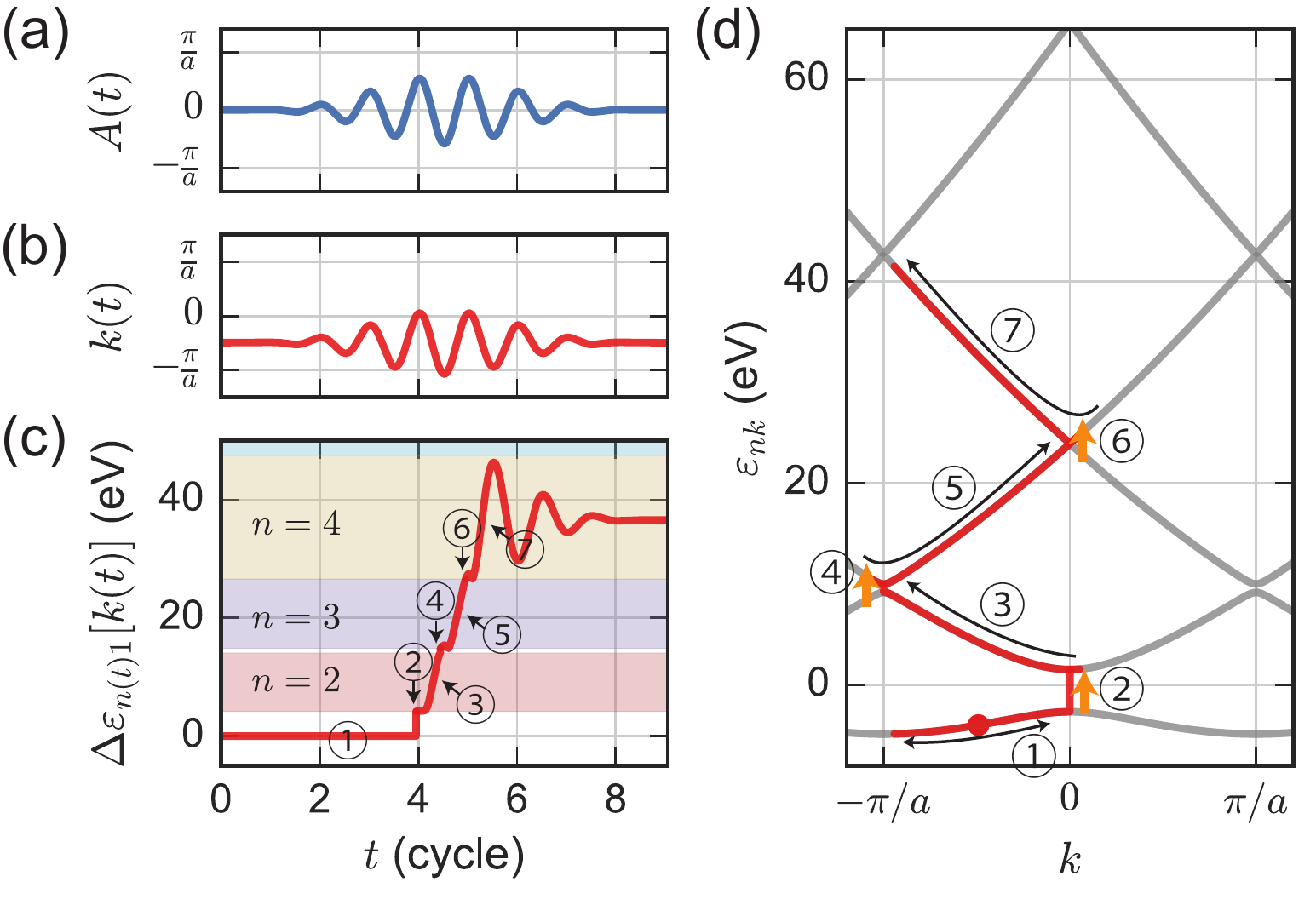}
 \caption{
 Momentum-space trajectory of an electron excited from a VB ($n=1$) to the first CB ($n=2$) at $t=4T$ ($T$ denotes an optical cycle), drawn based on the solid-state three-step model when $E_0 = 1.65$ V/nm or $\Apeak = 0.44 > \frac{\pi}{a}$, for which $k_0 = -0.49 \times \frac{\pi}{a}$.
 (a) waveform of $A(t)$
 (b) instantaneous crystal momentum $k(t)$
 (c) temporal evolution of the particle-hole energy, i.e., emitted photon energy
 (d) pictorial representation of the momentum-space electron trajectory in the band diagram.
 Reprinted with permission from Ref.~\cite{Ikemachi2017PRA}. Copyright 2017 by American Physical Society.
 }
 \label{fig:pictorial-strong}
\end{figure}

In analogy to the trajectory analysis of the gas-phase three-step model, explaining the cutoff law and the time-frequency structure, 
we can understand many aspects of solid-state HHG by using the above-mentioned recipes to trace electron trajectories in the band diagram.
An example for $\Apeak = 0.44 > \frac{\pi}{a}$ is displayed in Fig.~\ref{fig:pictorial-strong}.

Electrons initially in the valence band are accelerated (\textcircled{\scriptsize 1}),
and excited to the CB at $k=0$ at $t=t_0$ (\textcircled{\scriptsize 2}).
Vertical tunneling and recombination being assumed, once a waveform of $A(t)$ is given [Fig.~\ref{fig:pictorial-strong}(a)], the crystal momentum history is fully described as,
\begin{equation}
 k(t) = k_0 + A(t) = A(t) - A(t_0), \label{eq:momentum change in CB}
\end{equation}
regardless of the band where the electron resides [Fig.~\ref{fig:pictorial-strong}(b)].
It should especially be noticed that $|k(t)| < \Apeak$.
Thus, if $\Apeak < \frac{\pi}{a}$, the electron cannot reach the BZ edge but oscillates in the first CB without further excitation.
Therefore, the emitted photon energy is given by $\Delta\varepsilon_{21}[k(t)]$ as a function of recombination time $t$, and the cutoff energy is given by $\Delta \varepsilon_{21}(\Apeak)$, which agrees with the position represented by the white solid line in Fig. \ref{fig:harmonic spectrum}(b).

Now that $\Apeak > \frac{\pi}{a}$ in Fig. \ref{fig:pictorial-strong}, after excitation to the first CB (\textcircled{\scriptsize 1} - \textcircled{\scriptsize 2}), part of electrons can be accelerated to reach the BZ edge (\textcircled{\scriptsize 3}),
and open a channel to climb up to the upper CB (\textcircled{\scriptsize 4}) within a half cycle.
The promoted electrons then undergo intraband displacement to the reversed direction in the second CB ($n=3$) in the next half cycle, enabling photon emission of higher energy (\textcircled{\scriptsize 5}). 
This simple pictorial analysis neatly explains why multiple plateaus suddenly appear at $\Apeak \approx \frac{\pi}{a}$ [Fig.~\ref{fig:harmonic spectrum}(b)].
Electrons can experience interband transitions not only precisely at the minimum band gaps but also in their vicinities.
This is the origin of some high-energy components, which appear even before $\Apeak$ reaches $\frac{\pi}{a}$ in Fig.~\ref{fig:harmonic spectrum}(b), from $E_0 \sim 1.1$ V/nm.

Each time the electrons reach the minimum energy gap to next CB every half cycle, they either undergo further interband excitation (\textcircled{\scriptsize 5} - \textcircled{\scriptsize 7}) or pass through it.
They can climb up to the third CB ($n=4$) if $\Apeak^\prime < \frac{\pi}{a}$, where $\Apeak^{\prime}$ denotes the second maximum peak-to-valley amplitude (Fig. \ref{fig:field and current}), and the fourth CB ($n=5$) if $\Apeak^\prime > \frac{\pi}{a}$ at $t\approx 5.5 T$ with $T$ being the optical cycle.
From this scenario, we can estimate the maximum energy gain as
\begin{align}
 E_{c} = \left\{
\begin{array}{lc}	
\Delta \varepsilon_{41}(\Apeak^{\prime}) & (\Apeak^{\prime} < \frac{\pi}{a}) \\
\Delta \varepsilon_{51}(\Apeak^{\prime}) & (\frac{\pi}{a} < \Apeak^{\prime}),
\end{array}
 \right.
 \label{144641_16Nov16}
\end{align}
which reproduces the highest harmonic energy in Fig.~\ref{fig:harmonic spectrum}(b).
Thanks to the band-climbing process \footnote{This somewhat reminds us of {\it Donkey Kong}, an arcade game released by Nintendo (\url{https://en.wikipedia.org/wiki/Donkey_Kong_(video_game)})}, the highest cutoff energy can exceed the value expected in the gas phase for the same laser parameters and ionization potential (band gap energy in the solid case) \cite{Ndabashimiye_2016}.

Electrons that start from $k_0 \sim 0$ are excited when $A(t) \approx 0$, i.e., at an extremum of $E(t)$, promoting tunneling transition.
However, they cannot reach the BZ edge and are confined in the first CB if $\Apeak < \frac{2\pi}{a}$.
Consequently, their contributions are limited to the range below $E_{31}$.
In contrast, the harmonic components above $E_{31}$ including the highest cutoff are dominated by the electrons initially far from the $\Gamma$ point and
first excited in the vicinity of a peak of $A(t)$, where the electric field is weak, thus with smaller probability.
This may be one of the reasons why higher plateaus are weaker in intensity.

\subsection{Electron-Hole Interaction Effects}
\label{subsec:Electron-Hole Interaction Effects}

Section \ref{sec:Graphene} and Subsec.~\ref{subsec:Independent Electron Approximation} as well as most of the works investigating the mechanisms of solid-state HHG have used independent-electron approximation.
On the other hand, multielectron effects in the strong-field regime is largely unexplored.
Let us focus on the role of the electron-hole interaction (EHI), which forms excitions in the linear response regime, in this Subsection, based on the time-dependent Hartree-Fock (TDHF) calculation \cite{Ikemachi2018PRA}.

We again consider a 1D model crystal along laser polarization.
A 1D system, which has a strong electron-hole correlation \cite{haug2009quantum}, is suitable for the investigation of EHI.
We solve a set of the spin-restricted TDHF equation,
\begin{align}
 i \frac{\partial}{\partial t} \psi_{nk_0}(x, t) = \hat{h}(t) \psi_{nk_0}(x, t)
 = \left[ \frac{1}{2}[\hat{p} + A(t)]^2 + U(x) + \hat{w}[\rho(t)]\right] \psi_{nk_0}(x, t),\label{eq:TDHF equation}
\end{align}
in the velocity gauge, where $U(x)$ denotes the periodic potential from the crystal nuclei, $\rho(t)$ the density matrix,
\begin{equation}
 \rho (x, x^{\prime}, t) = 2\sum_{n\in {\rm VB}, \; {k_0}} \psi_{nk_0}(x,t)\, \psi_{nk_0}(x^{\prime},t)^*, \label{eq:density matrix operator}
\end{equation}
and the operator $\hat{w}[\rho]$, composed of the Coulomb and exchange terms, describes the contribution from the interelectronic Coulomb interactions, reflecting the dynamics of the other electrons within a mean-field treatment.
As the initial state of $\psi_{nk_0}(t)$, we take the VB Bloch function $\phi_{nk_0}$, obtained as the self-consistent eigenstate of the field-free Hartree-Fock Hamiltonian
with the energy eigenvalue $\varepsilon_{b{k_0}}$.
We calculate the HHG spectrum as the modulus square of the Fourier transform of the induced current $j(t) = 2\sum_{n\in{\rm VB}, \; {k_0}} \bra{\psi_{nk_0}(t)} \hat{p} + A(t) \ket{\psi_{nk_0}(t)}$.

Let us compare the TDHF equation Eq.~(\ref{eq:TDHF equation}) with the independent-electron TDSE Eq.~(\ref{eq:TDSE in general}).
Aside from the exchange terms not included in the latter, the effective potential $V(x)$ is considered to include the Coulomb terms formed by the initial state, in addition to $U(x)$.
Therefore, to mimic the independent-electron treatment, we also perform simulations using the {\it frozen} TDHF Hamiltonian
\begin{equation}
 \hat{h}_f(t) = [\hat{p} + A(t)]^2 / 2 + U(x) + \hat{w}[\rho_0],
\end{equation}
with $\rho_0(x, x^{\prime}) = e^{-iA(t) \cdot x} \rho(0) e^{iA(t) \cdot {x'}}$, where electrons move independently in the potential constructed by the ground state Bloch functions.
The factors $e^{-iA(t)\cdot x}$ and $e^{iA(t) \cdot x'}$ are required in the velocity gauge.
The difference $\hat{w}[\delta\hat{\rho}(t)]$ with $\delta\hat{\rho}(t) = \hat{\rho}(t) - \hat{\rho}_0$ between the {\it full} TDHF Hamiltonian $\hat{h}(t)$ and frozen Hamiltonian $\hat{h}_f(t)$ takes account of EHI.

Specifically, our system is a 1D model hydrogen chain insulator with a lattice constant of $a = 3.6$ a.u., composed of a series of hydrogen dimers whose bond length is $1.6$ a.u..
We use a soft-Coulomb potential $v(x, x^{\prime}) = [(x-x^{\prime})^2 + 1]^{-1/2}$ for both electron-nucleus and electron-electron interactions.
Figure \ref{fig:pictorical} shows the band structure, the set of the energy eigenvalues $\varepsilon_{nk_0}$, with a gap energy of $9.5$ eV.
The lowest band or VB is initially fully occupied.
The laser field is assumed to be $E(t) = E_0\sin^2(t/\tau)\sin(\omega t)$ with $\tau = 702.3 $ (5 cycle), $\hbar \omega = 0.387$ eV.
We numerically integrate the full and frozen TDHF equations, using the finite-difference method with the grid spacing $0.24$ atomic units, time step size $4.4 \times 10^{-3}$ atomic units, and the number of $k$ points $201$.

\begin{figure}[tb]
 \centering
 \includegraphics[width=1.0\linewidth]{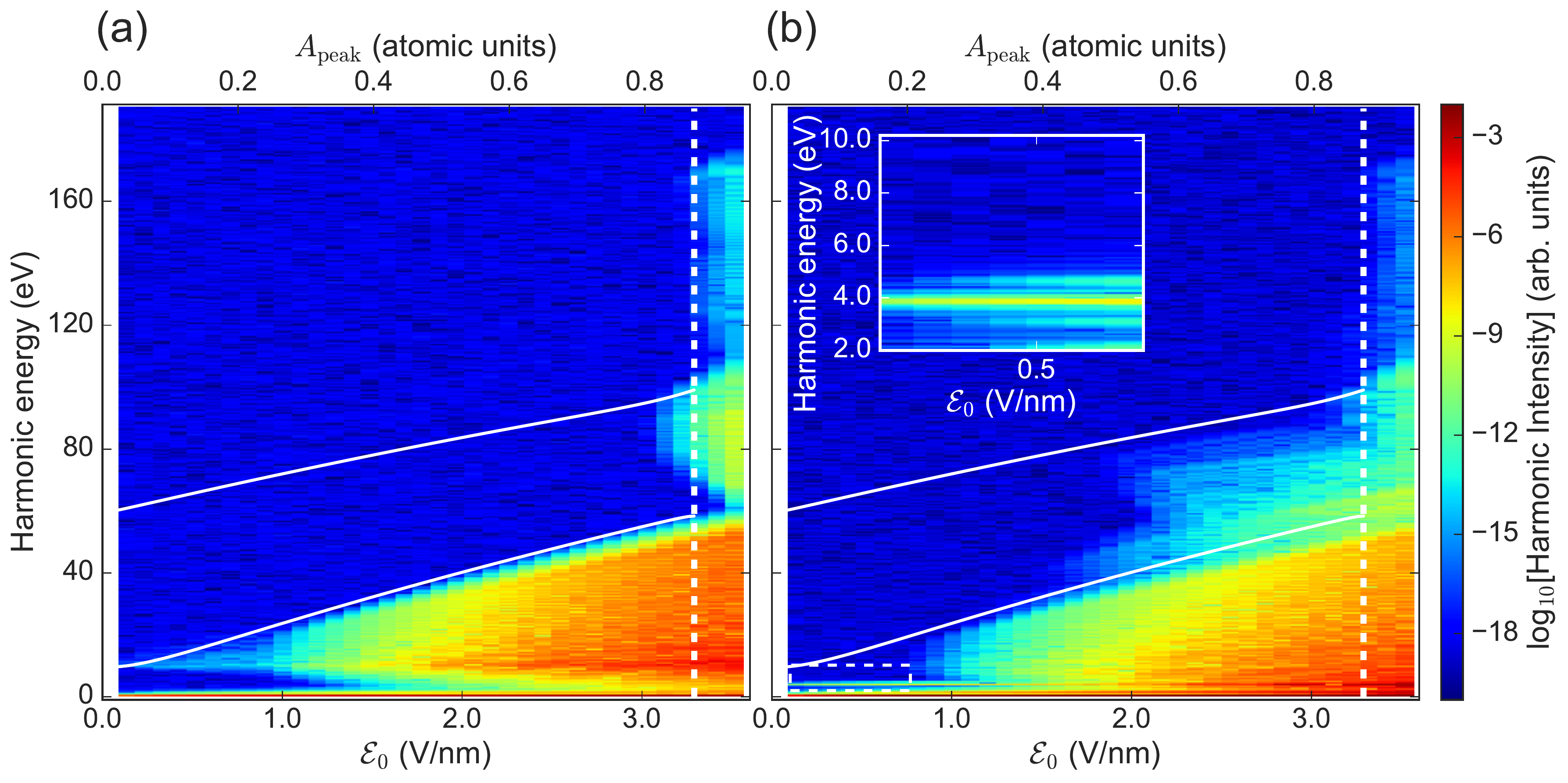}
 \caption{
 Harmonic spectra as functions of the field amplitude $\mathcal{E}_0$ (bottom axis) and corresponding $\Apeak$ (top axis) obtained from (a) frozen TDHF and (b) full TDHF simulations.
 The white dashed vertical lines denote $\Apeak = \frac{\pi}{a}=0.87$, which characterizes the position where the multiple plateaus appear according to the solid-state three-step model \cite{Ikemachi2017PRA}.
 Two white solid lines are the energy differences between CBs and VB as function of $\Apeak$, i.e., $\varepsilon_{10}(\frac{\pi}{a} - \Apeak)$ (lower) and $\varepsilon_{20}(\Apeak)$ (higher).
 Inset: close-up of the low-field region represented by a dashed rectangle in (b).
 Reprinted with permission from Ref.~\cite{Ikemachi2018PRA}. Copyright 2018 by American Physical Society.
 }
 \label{fig:harmonic spectra}
\end{figure}

\begin{figure}[tb]
 \centering
 \includegraphics[width=1.0\linewidth]{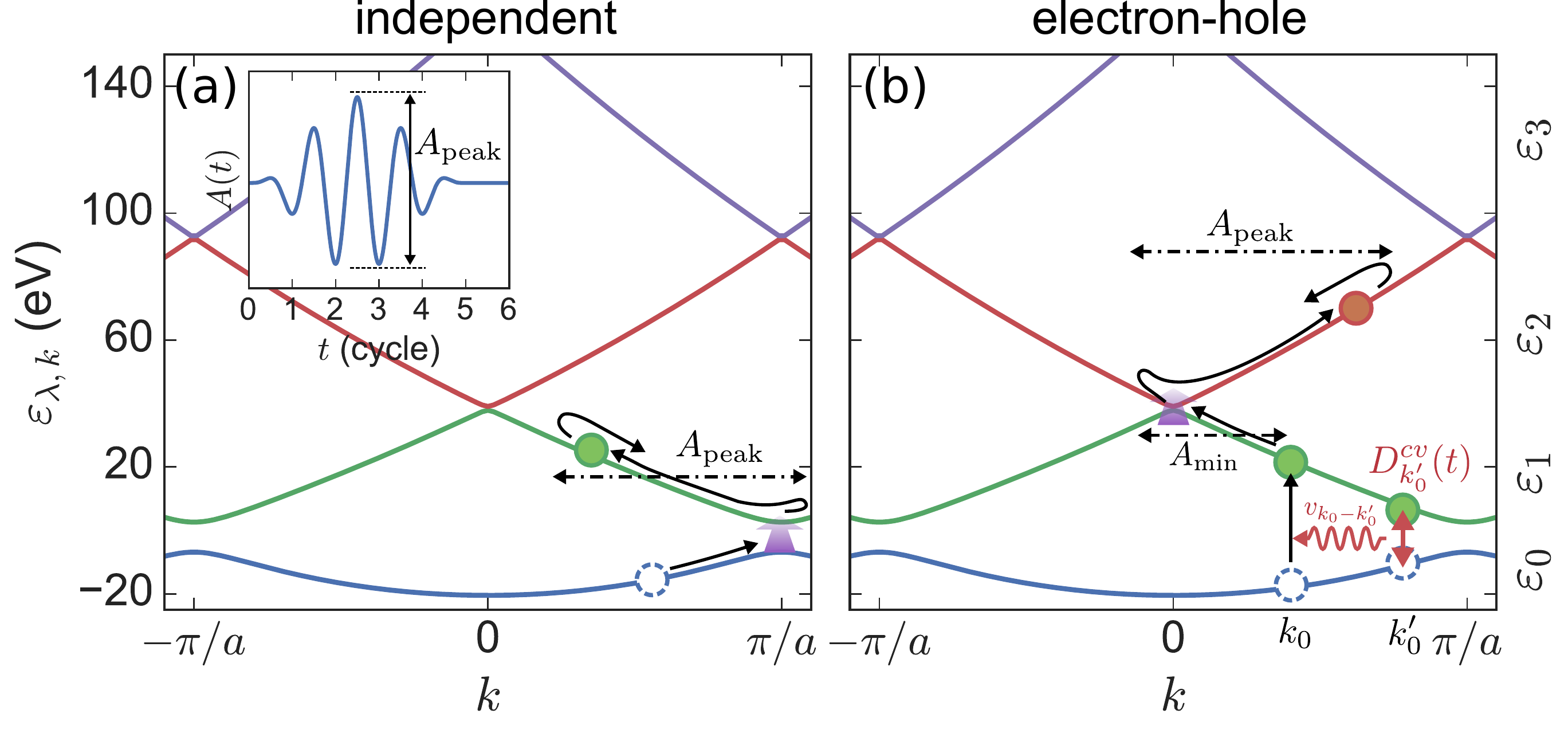}
 \caption{
 Pictorial representation of momentum-space electron dynamics (a) within the independent-electron approximation and (b) involving hauling-up excitation.
 The inset in (a) shows the waveform of the vector potential used in TDHF and {\it frozen} TDHF simulations and the definition of $\Apeak$.
 The single VB and first three CBs are shown for a 1D model hydrogen chain insulator (see text).
 The band index $n$ is labeled as $0, 1, 2, \dots$ from the bottom.
 Reprinted with permission from Ref.~\cite{Ikemachi2018PRA}. Copyright 2018 by American Physical Society.
 }
 \label{fig:pictorical}
\end{figure}

Figure \ref{fig:harmonic spectra} displays the calculated harmonic spectra as functions of the field amplitude $E_0$ and the corresponding $\Apeak$ [inset of Fig.~\ref{fig:pictorical}(a)] 
In the case of the {\it frozen} TDHF [Fig.~\ref{fig:harmonic spectra}(a)], i.e., within the independent-electron approximation, we can well understand the appearance of multiple plateaus at $\Apeak = \frac{\pi}{a} = 0.87$ and the cutoff positions on the basis of the solid-state momentum-space three-step model \cite{Ikemachi2017PRA,Wu2016PRA,Du2017OE} discussed in Sect. \ref{subsec:Independent Electron Approximation}.
A typical trajectory is depicted in Fig.~\ref{fig:pictorical}(a) for $\Apeak < \frac{\pi}{a}$, for which no excited electrons in the first conduction band (CB) can reach the next MBG ($k=0$), and they only oscillate in the first CB, which forms a single plateau in the high-harmonic spectra.

The full TDHF results, with EHI turned on, are shown in Fig.~\ref{fig:harmonic spectra}(b).
We find two distinct features.
First, at low intensity [inset in Fig.~\ref{fig:harmonic spectra}(b)], there is an exciton peak at $3.8$ eV below the gap energy, which indicates that the TDHF simulations capture EHI appropriately.
Note that TDDFT at present cannot reproduce excitons, which is based on the simple adiabatic local-density approximation in practical implementations without any nonlocal exchange-like term \cite{Onida2002}.
\footnote{The excitonic physics seems to be guaranteed by the mixture of the exchange term, called hybrid-functional, within the TDDFT framework \cite{Paier2008, Sato2015, Penmmaraju2019}. However, it is not fully investigated that the potential of the hybrid-functional for electron excitation of the extended systems due to few applications.}
Second and more remarkably, the second plateau already appears at $\Apeak \sim 0.5$, much smaller than $\frac{\pi}{a}$.
Thus, EHI qualitatively alters HHG spectra.

In order to understand the microscopic mechanism underlying the latter feature, let us expand the orbital functions $\psi_{b k_0}(x,t)$ with Houston states $e^{-i A(t) x} \phi_{nk(t)}(x)$ \cite{Krieger1986}, the instantaneous eigenstates of $\hat{h}_f(t)$ with eigenvalues $\varepsilon_{nk(t)}$, as
\begin{equation}
 \psi_{nk_0}(x,t) = \sum_m \alpha_{nk_0}^{m}(t) e^{-i \int_0^t \varepsilon_{m k(t^{\prime})} dt^{\prime}}e^{-i A(t) x} \phi_{mk(t)}(x),\label{eq:Houston function expansion of orbital}
\end{equation}
where $k(t) = k_0 + A(t)$ is the instantaneous crystal momentum incorporating intraband dynamics.
Since the system under consideration has a single VB, we drop the initial band index $n$ hereafter.
Substituting Eq.~(\ref{eq:Houston function expansion of orbital}) into Eq.~(\ref{eq:TDHF equation}), we obtain equations of motion for complex amplitudes $\alpha_{k_0}^m(t)$ expressing interband dynamics,
\begin{equation}
 i\frac{d}{dt} \alpha_{k_0}^m(t) = \sum_{n} \alpha_{k_0}^{n}(t) e^{i\int_0^t \Delta\varepsilon_{mn}[k(t^{\prime})]dt^{\prime}} \left( E(t) d_{k(t)}^{mn} + \bra{\tilde{\phi}_{mk_0}} \hat{w}[\delta\rho(t)] \ket{\tilde{\phi}_{nk_0}}\right),\label{eq:EOS for Houston probability amplitude}
\end{equation}
where $d_{k}^{mn} = i\braket{u_{km} | \nabla_k u_{kn}}$ with $u_{km}(x)$ being the lattice periodic part of the initial Bloch state, i.e., $\phi_{km}(x) = e^{i k x} u_{km}(x)$, and $\tilde{\phi}_{nk_0}(x,t)=e^{-iA(t)x}\phi_{nk(t)}(x)$. 
The first term comes from the {\it frozen} TDHF Hamiltonian, and thus describes the independent electron dynamics.
The second term, on the other hand, stems from EHI $\hat{w}[\delta\rho(t)]$.
After some approximation and algebraic manipulations \cite{Ikemachi2018PRA}, we get,
\begin{align}
 i\frac{d}{dt} \alpha_{k_0}^m(t) = \sum_{n} \alpha_{k_0}^{n}(t) e^{i\int_0^t \Delta\varepsilon_{mn}[k(t^{\prime})]dt^{\prime}} 
  \left[ E(t) d_{k(t)}^{mn} - \sum_{q \in {\rm BZ}} \bar{v}(-q) D_{k(t)+q}^{mn}(t) \right],\label{eq:EOM for Houston function expansion}
\end{align}
where $\bar{v}(q)$ denotes the spatial Fourier transform of the interelectronic soft Coulomb potential, and $D^{mn}_{k(t)}$ the time-dependent interband polarization between $m$ and $n$ at $k(t)$:
\begin{equation}
 D^{mn}_{k(t)}(t) = \alpha_{k_0}^m(t) \alpha_{k_0}^{n\ast}(t) e^{-i \int_0^t \Delta\varepsilon_{mn}[k(t^{\prime})] dt^{\prime}}.
\end{equation}

Since the population of CBs turns out to be small ($\lesssim 10^{-3}$) \cite{Ikemachi2018PRA}, we introduce approximations $\alpha_{k_0}^{0}(t) \approx 1$ and $\alpha_{k_0}^{m \ge 1}(t) \approx 0$ \cite{McDonald2017}.
Then Eq.~(\ref{eq:EOM for Houston function expansion}) for the first CB ($m=1$) becomes
\begin{equation}
 i\frac{d}{dt} \alpha_{k_0}^1(t) \approx e^{i\int_0^t \varepsilon_{10}[k(t^{\prime})]dt^{\prime}} 
   \left[ E(t) d_{k(t)}^{10} - \sum_q \bar{v}(-q) D_{k(t)+q}^{10}(t) \right],
   \label{eq:EOM for Houston with FVB}
\end{equation}
for the excitation dynamics of a VB electron with an initial crystal momentum $k_0$.
The second term due to EHI indicates that interband or electron-hole polarization at a remote crystal momentum $k(t) + q$,
\begin{equation}
 D_{k(t)+q}^{10}(t) = \alpha_{k_0+q}^1(t) e^{-i\int_0^t \varepsilon_{10}[k(t^{\prime})+q]dt^{\prime}},
\end{equation}
can induce quasi-resonant excitation when $\varepsilon_{10}[k(t)] \approx \varepsilon_{10}[k(t)+q]$.
Therefore, even if a VB electron starting from $k_0$ dose not reach MBG through intraband displacement, it can be excited to the first CB once another electron initially at $k_0+q$ reaches MBG and tunnels to the CB [Fig.~\ref{fig:pictorical}(b)].
It should be noticed that neither the first nor second terms directly change the crystal momentum, thus, the instantaneous crystal momentum is always given by $k(t) = k_0 + A(t)$, in whichever band the electron actually is. 

This {\it hauling-up} effect provides a shortcut for VB electrons to climb up to the second CB, which leads to the formation of the second plateau even if $\Apeak < \frac{\pi}{a}$.
The electrons initially at $k_0 \in [-\max(A(t)), -\min(A(t))]$ pass by $k=0$, i.e., MBG between the first and second CB.
Thus, if these VB electrons are excited to the first CB via the hauling-up effect, then they can climb up to the second CB by tunneling at $k=0$, eventually forming the second plateau via recombination with the VB hole.
Note that they cannot reach MBG at $k=\pm\frac{\pi}{a}$ between the second and third CB.
Therefore, the cutoff energy is expected to be given by $\varepsilon_{20}(\Apeak)$.
This prediction is in good agreement with the cutoff energy obtained from the TDHF simulation at $0.5 \lesssim \Apeak \le \frac{\pi}{a}=0.87$ [the upper white line in Fig.~\ref{fig:harmonic spectra}(b)].

\section{Time-Dependent Density-Matrix Method Combined with First-Principles Calculation for Three-Dimensional Crystals}
\label{sec:TDDM}

In material science, density-functional theory (DFT) is one of the {\it de facto} standards for materials at electronic ground state, owing to a good balance between accuracy and computational cost. Time-dependent density-functional theory (TDDFT) is one of the most feasible theories to describe electron excitation under an intense laser field from first-principles \cite{Otobe2012, Klemke2019}. 
While TDDFT shows accurate results, its calculation cost is relatively expensive, e.g. a few hundred core-hour or longer for well-converged results of a laser parameter.
This calculation cost is still high for modern supercomputers when we need to investigate the optical response over a wide parameter region for the laser pulse, such as photon energies, field strengths, polarization properties, and so on.
We develop an alternative theoretical framework based on a first-principles theory with cheaper calculation cost, called time-dependent density-matrix (TD-DM) method.

A one-body density-matrix (DM) is the key degree of freedom in TD-DM. The equation of motion is von Neumann equation:
\begin{align}
i\frac{d\rho}{dt}
=
\left[h(t),\rho\right]
+
\left(i \frac{d\rho}{dt}\right)_{\mathrm{coll}}\label{eq:TD-DM_EOM}
\end{align}
where $h$ and $\left(i d\rho/dt\right)_{\mathrm{coll}}$ are one-body Hamiltonian and collision term.
This is a standard approach in the nonlinear optics \cite{Boyd}. Our TD-DM can be a first-principles theoretical framework by choosing a representation that orbitals derived from self-consistent DFT.
The matrix elements of DM, the Hamiltonian, and the collision term are expressed by the orbitals $\phi_{b{\bf k}}$, where $b$ and ${\bf k}$ are indices for a band and Brillouin zone, respectively.
When the collision term is neglected, this theoretical framework is equivalent to independent electron dynamics for DFT one-body Hamiltonian.
This framework is regarded as a generalization of the one-dimensional theories in the precedent sections to a spatially three-dimensional system with the one-body potential $V({\bf r})$ obtained by the DFT calculation.
Our TD-DM includes the many-body effect of the electrons reflected in the constitution of the electronic structure through DFT calculation. 
In other words, the dynamical correction of the many-body effect is not included in our TD-DM, except for effect provided by the collision term.
The computational cost of this framework is substantially reduced by using a basis set, compared to grid-basis TDDFT \cite{SALMON, Octopus}.
An additional advantage of this framework is flexible modifications on the top of DFT, {\it e.g.} bandgap correction and phenomenological relaxations via the scattering term.
This theoretical framework is close to the de fact standard theory, semiconductor Bloch equation (SBE) \cite{Lindberg1988}. The difference lies in the length gauge and the electron-hole attraction term.

The key observable of our TD-DM is induced current density evaluated by the expectation of the velocity operator ${\bf v}(t)={\bf p}+{\bf A}(t)$ divided by the volume of cell $V_{\mathrm{cell}}$ : ${\bf J}(t)=-\mathrm{tr}\left(\rho {\bf v}\right)/V_{\mathrm{cell}}$.
To obtain emitted photon intensity, taking the absolute value of Fourier transform of the acceleration density, the temporal derivative of the current density. 

\begin{figure}[tb]
\centering
\includegraphics[width=1.0\linewidth]{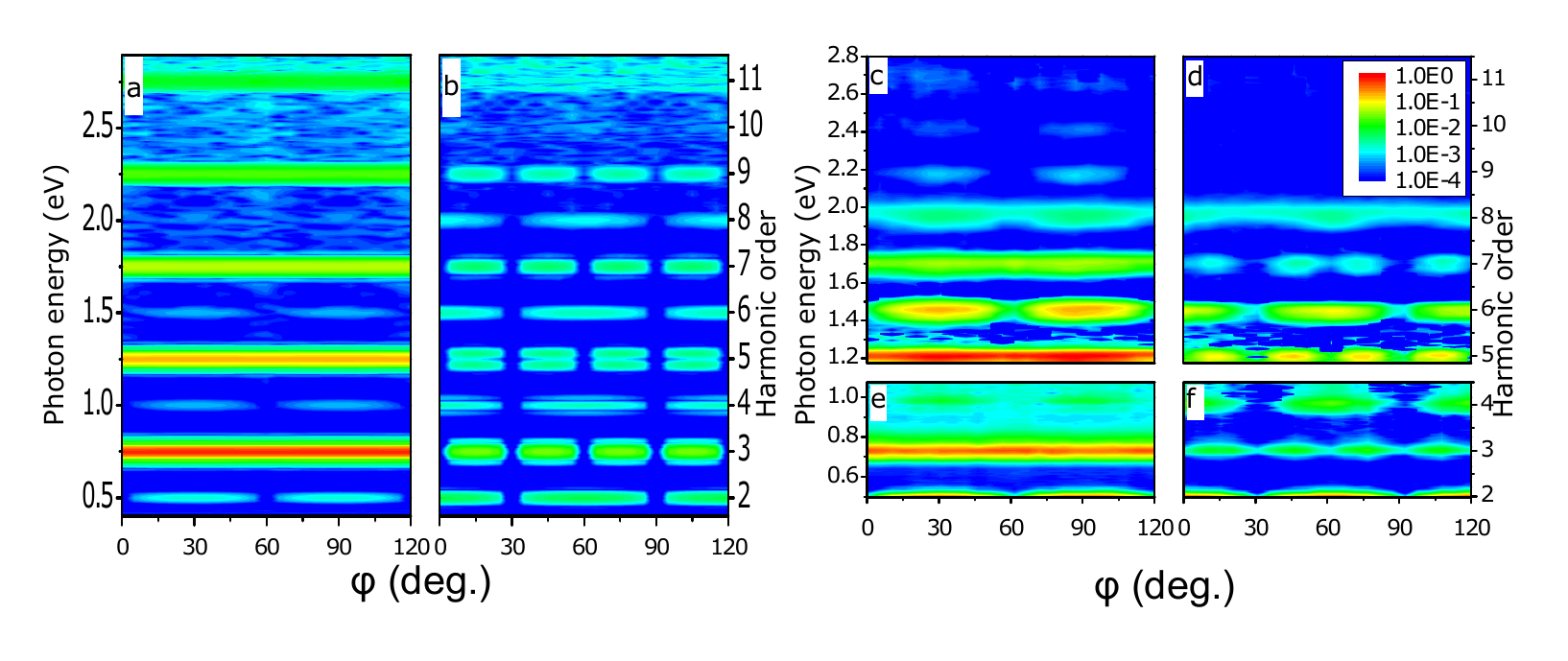}
\caption{Harmonic spectra from GaSe crystal. Parallel (a) and perpendicular (b) components from TD-DM simulation. (c) and (d) are the same as (a) and (b) but experimental results. Reprinted with permission from Ref.~\cite{Kaneshima2018}. Copyright 2018 by American Physical Society.}
\label{fig:HHG_GaSe}
\end{figure}

We show our TD-DM performance by an application to a high-harmonic generation from GaSe crystal \cite{Kaneshima2018}.
In the experiment, the GaSe sample is exposed to linearly polarized light whose polarization is on the basal plane, in which the wavelength and the pulse duration are 4.96 $\mu$m and 200 fs. 
Angle dependences of the driving field with respect to crystal orientation are measured for two orthogonal polarization directions of the emitted photon, parallel, and perpendicular components to the driving field.
The results are shown in Fig.~\ref{fig:HHG_GaSe} (a)-(d), together with the theoretical counterparts. 
Our TD-DM reproduces almost all features in the experimental spectra, almost isotropic angle dependence for the parallel component of odd-order harmonics, a 60-degree period of even-order harmonics, and a 30-degree period for the perpendicular components of odd-order harmonics.

  This direct comparison between the theoretical simulation and the experiment is hardly achieved if we employ TDDFT because of the tough computational cost.
  To achieve an expected feature for the polarization direction dependence, very dense Brillouin zone sampling, $64\times 64\times 12$, was mandatory.
  Besides, we must perform multiple calculations for different angles of the field polarization.
  Computations to draw Fig. \ref{fig:HHG_GaSe} (a)-(b) requires 80 thousands core-hour.
  Typically TDDFT requires tens to a hundred times more.
  The core-hour estimation for TDDFT is possible with modern supercomputers in principle but unrealistic for daily use of the supercomputers.
  Our TD-DM is a lightweight simulation option to perform a more comprehensive investigation of strong-field phenomena in solids, keeping the nonempirical nature.

The polarization-resolved analysis showed that crystal symmetry is reflected in the HHG spectra even for the non-perturbative regime beyond the susceptibility-based argument for the second- and third-order harmonics. 
Part of the fingerprint of HHG for the symmetry can be understood by the intraband current model with a time-independent carrier population \cite{Ghimire2011, You2017} capturing band-structure anisotropy of crystals. 
While this intraband current model gives us a clear-cutting and simple description of the symmetric aspects, a qualitative judgment of the intraband current is severe because of many assumptions to proceed with the model calculations. 
We made a scheme to decompose the current density, in our TD-DM, into intraband and interband contributions, like the intraband current and the interband polarization in the SBE. 

\begin{figure}[tb]
\centering
\includegraphics[width=1.0\linewidth]{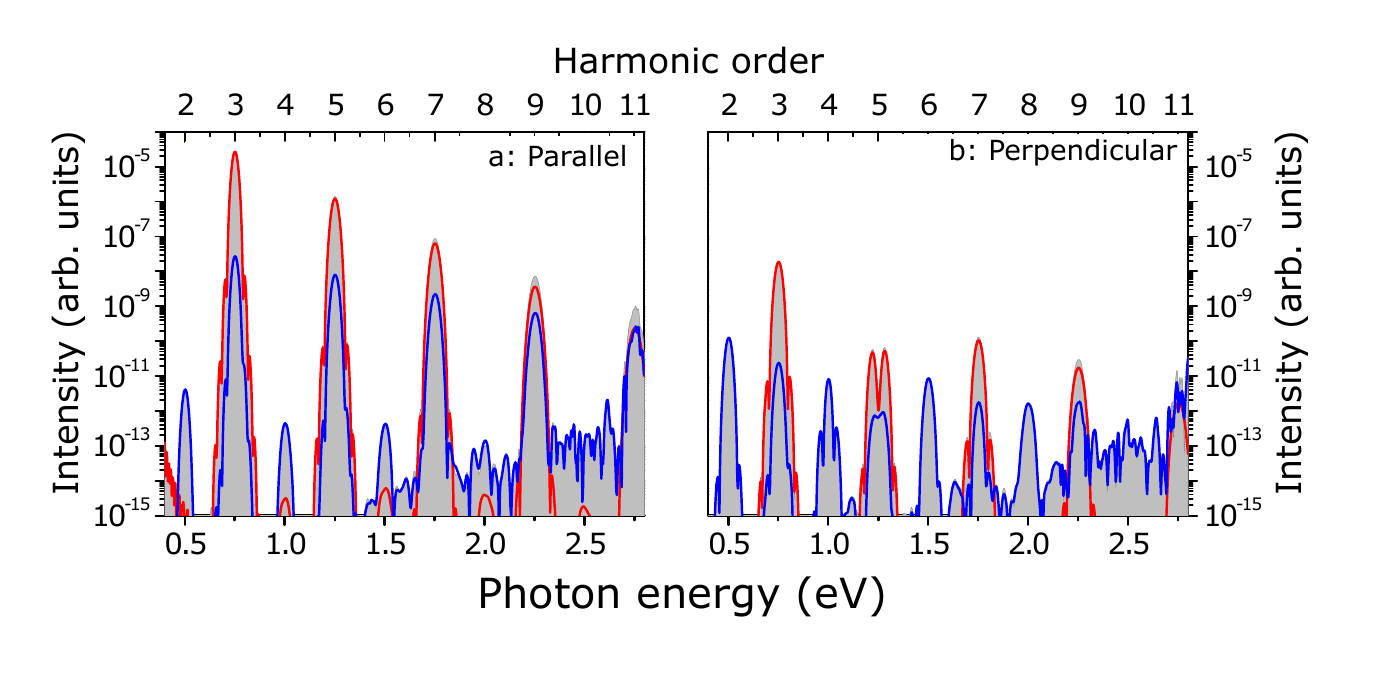}
\caption{Decompositions of harmonic spectra into intraband and interband contributions, for parallel (h) and perpendicular (i) components. Total (gray shadow), intraband (red solid), and interband (blue solid) components are shown. Reprinted with permission from Ref.~\cite{Kaneshima2018}. Copyright 2018 by American Physical Society.}
\label{fig:intra-inter}
\end{figure}

We define the intraband component of the current density as a partial sum of the trace over only the diagonal contribution of DM represented by the instantaneous eigenfunction of the time-dependent Hamiltonian $h(t) \varphi_{\beta k}^{(t)} = \varepsilon_{\beta k}^{(t)}\varphi_{\beta k}^{(t)}$ as
\begin{align}
{\bf J}(t) &= {\bf J}_{\mathrm{intra}}(t) + {\bf J}_{\mathrm{inter}}(t),\label{eq:intra+inter} \\
{\bf J}_{\mathrm{intra}}(t)
&=
-\frac{1}{V_{\mathrm{cell}}}\sum_{\beta {\bf k}} \left\langle \varphi_{\beta k}^{(t)} \left| \rho \right| \varphi_{\beta k}^{(t)} \right\rangle \left\langle \varphi_{\beta k}^{(t)} \left| {\bf v} \right| \varphi_{\beta k}^{(t)} \right\rangle,\label{eq:intra} \\
{\bf J}_{\mathrm{inter}}(t)
&=
-\frac{1}{V_{\mathrm{cell}}}\sum_{\beta\gamma (\beta \neq \gamma) {\bf k}} \left\langle \varphi_{\beta k}^{(t)} \left| \rho \right| \varphi_{\gamma k}^{(t)} \right\rangle \left\langle \varphi_{\gamma k}^{(t)} \left| {\bf v} \right| \varphi_{\beta k}^{(t)} \right\rangle \label{eq:inter}
\end{align}
where the interband contribution is obtained as the rest of the total current subtracted by the intraband contribution or the partial sum of the trace over the off-diagonal component of the DM. 
The superscript parenthesis $t$ of the variables, $\bullet^{(t)}$, represent that the object parametrically depends on the time. 
The intraband contribution defined here is the sum over product between the population evaluated with the instantaneous basis and the group velocity of the band structure because the velocity expectation evaluated with the instantaneous Hamiltonian is equivalent to the group velocity with the vector potential $\left\langle \varphi_{\beta k}^{(t)} \left| {\bf v}(t) \right| \varphi_{\beta k}^{(t)} \right\rangle = \left(\partial \varepsilon_{\beta {\bf k}} /\partial {\bf k}\right)_{{\bf k}+{\bf A}(t)}$. 
This formula is nicely related to the intraband current in SBE. 
This definition of the intraband current is a generalization of the simplest intraband current model, such that the time-dependent population obtained from the microscopic theory rather than just constant.
The harmonic spectra of the two contributions are shown in Fig.~\ref{fig:intra-inter}. 
In the lower odd-order harmonics, the intraband contribution dominates the total yield of the harmonics. 
The interband contributions increase with increasing the photon energy toward 2 eV, which is the bandgap of the GaSe. 
The two contributions become comparable for the ninth and eleventh harmonics.

One of the most striking results is that all even-order harmonics are exclusively from the interband contributions. This fact invokes that band structure with spin-independent Hamiltonian is always spatially symmetric. 
The intraband current is expected not to produce even-order harmonics because of the symmetric band structure. 
A significance of this investigation is that the expected results are demonstrated by a microscopic quantum mechanical simulation based on a first-principles theory. 
The exclusion of the even-harmonic for the intraband current supports our definition of the intraband and interband contributions to the total current density.

The intraband current can be further decomposed into band-resolved contributions by taking the partial sum of Eq. \eqref{eq:intra} over a part of bands. 
This analysis allows us to investigate which bands, associated with atomic orbital nature, mainly produce the harmonics for a specific situation. 
We apply this analysis to HHG from the $\mathrm{CsPbCl}_3$ perovskite \cite{Hirori2019}.

\begin{figure}[tb]
\centering
\includegraphics[width=0.8\linewidth]{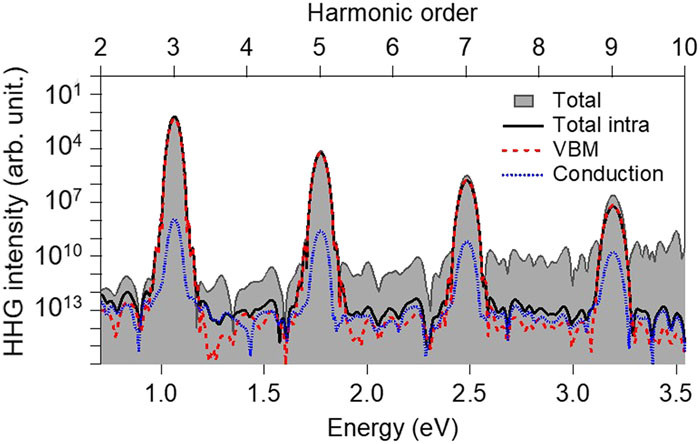}
\caption{ Decompositions of harmonic spectra into band-resolved intraband and interband contributions. Total (gray shadow), total intraband (black solid), VBM intraband (red dashed), and conduction intraband (blue solid) components are shown. Reprinted from Ref.~\cite{Hirori2019} with permission under the Creative Commons Attribution (CC BY) license.}
\label{fig:band-resolved_intra}
\end{figure}

We show the band-resolved intraband contribution to HHG in Fig.~\ref{fig:band-resolved_intra}. 
We employ a pulsed electric field that has 0.62 eV photon energy, 1.0 V/nm field strength, 160 fs full width at half maximum pulse duration. 
Reflecting on the inversion symmetry of the crystal, only odd-order harmonics appear in the spectrum.
The total intraband current dominates the power spectrum of the total current density. 
For a perovskite containing halide and lead ions, the topmost valence bands have the characteristic function for the optical absorption and are frequently argued as valence band maximum (VBM). 
VBM of the halide-lead perovskite is composed of p-nature halide and s-nature leads orbitals. 
According to the band-resolved intraband current analysis, we define VBM and conduction intraband currents as partial sums of the Eq. \eqref{eq:intra} over the VBM and conduction bands:
\begin{align}
{\bf J}_{\mathrm{VBM}}(t)
&=
-\frac{1}{V_{\mathrm{cell}}}\sum_{\beta (\in \mathrm{VBM}) {\bf k}} \left\langle \varphi_{\beta k}^{(t)} \left| \rho \right| \varphi_{\beta k}^{(t)} \right\rangle \left\langle \varphi_{\beta k}^{(t)} \left| {\bf v} \right| \varphi_{\beta k}^{(t)} \right\rangle, \\
{\bf J}_{\mathrm{conduction}}(t)
&=
-\frac{1}{V_{\mathrm{cell}}}\sum_{\beta (\in \mathrm{conduction}) {\bf k}} \left\langle \varphi_{\beta k}^{(t)} \left| \rho \right| \varphi_{\beta k}^{(t)} \right\rangle \left\langle \varphi_{\beta k}^{(t)} \left| {\bf v} \right| \varphi_{\beta k}^{(t)} \right\rangle \label{eq:VBM-conduction}
\end{align}
The contributions of the two intraband currents are shown as a red dashed line in Fig.~\ref{fig:band-resolved_intra}. 
The intensity of the total intraband current is well dominated by the VBM components.
The conduction band does not affect almost anything for the spectrum. 
The conduction band is frequently regarded as a source of the intraband band harmonic generation within the simplest intraband current model, because of more dispersive band curves than valence ones. 
This explicit decomposition raises a counter-intuitive point that VBM intraband current due to the hole motion mainly produces the HHG, at least for $\mathrm{CsPbCl}_3$ perovskite.

\section{Dynamical Franz-Keldysh Effect}
\label{sec:DFKE}
\label{sec:1}
In recent years, it has become possible to generate attosecond pulsed light using high-order harmonics, which are nonlinear interactions between the gas phase and ultrashort pulse lasers.
Currently, the pulse width is reduced to tens of attoseconds.
Since the 2010s, changes in optical properties of solids shorter than the electric field period of light have been reported.

Phenomena that occur inside a solid in a laser field can be broadly classified according to the presence or absence of electronic excitation. 
Since the recombination of electron-hole pairs is generally on a long time scale of the order of pico-second, electronic excitation causes a change in physical properties that is not recovered in the attosecond time scale.
Many phenomena including the electron excitation, such as the saturable absorption due to occupation of the conduction band, and generation of higher harmonics in solids have been attracting interests. 
On the other hand, it is considered that the phenomenon that recovers quickly after passing through the laser does not contribute to electronic excitation.

A strong laser can be treat as an oscillating electric field. 
The change in the dielectric function without electronic excitation by an electrostatic field is known as the Franz-Keldysh effect (FKE).
FKE is photon absorption via the tunnel effect and can be considered as a long wavelength or strong laser field limit. 
The dynamical Franz-Keldysh effect (DFKE) is an extension of FKE to periodic oscillating electric fields \cite{Jauho96}.
Although DFKE has been investigated theoretically and experimentally for the time-averaged modulation which corresponds to the blueshift 
of the band gap by the ponderomotive  energy \cite{Jauho96,Jauho98,Nordstrom98}. 
Time-resolved observation of the DFKE is reported by Novelli {\it et al.}, employing the THz light \cite{Novelli}. 
We have recently proposed the time-resolved analysis for dynamical Franz-Keldysh effect (Tr-DFKE).
Analytical theory and first-principles simulations have revealed that Tr-DFKE is a modulation faster than the oscillation of the laser electric field.
Recently, the sub-cycle modulation has been confirmed experimentally.
In this Section, we would like to introduce our recent works on the construction of DFKE analytical theoretical formulas and first-principles calculations.

\subsection{Time-resolved spectroscopy}
\label{sec:2}

We would like to clarify time-resolved spectroscopy before moving on to specific processe.
The optical property of a material is described by susceptibility $\chi$, 
which connects the polarization $P (t)$ to a given electric field E $(t)$ as
$P_{i}(t)=\sum_{j}\int_{-\infty}^{\infty}dt'\chi_{ij} (t-t')E_{j}(t') $.
Here indexes $i$ and $j$ indicate the components $(x,y,z)$ of the vector.
It should be noted that the $\chi$ is the function of the relative time, $t-t'$.
The time-dependence of $\chi$ indicates the time-invariance of the optical properties.
If the system depends on the time, we should reconsider the $\chi$ as the function of two independent time,
\begin{equation}
\chi(t-t')\rightarrow\chi'(t,t').
\end{equation}
In the same way, dielectric function $\varepsilon$ and optical conductivity $\sigma$ become the function of $t$ and $t'$,
$\varepsilon_{ij}(t,t')=\delta_{ij}\delta(t-t')+4\pi \chi'_{ij}(t,t'),
\chi'_{ij}(t,t')=\int_{-\infty}^{\infty} dt" \Theta(t-t")\sigma_{ij}(t",t')$,
where $\Theta(t-t")$ is the Heaviside function.

The optical properties is observed as the modulation of the probe pulse whose peak intensity is at the time $T_p$.
The detected susceptivity is the function of the frequency $\omega$ and the $T_p$, $\chi(T_p,\omega)$.
If we assume the probe pulse as $f_p\delta(t-T_p)$, the polarization becomes 
$P(t)=f_p\chi'(t,T_p)$.
In this step, we assume that the $\chi'(t,T_p)$ has only diagonal part.
The susceptivity in the frequency-domain $\chi(T_p,\omega)$ can be defined from  the $\chi'(t,T_p)$, 
$\chi(T_p,\omega)=\int dt e^{i\omega t}P(t)/\int dt e^{i\omega t} f_p\delta(t-T_p)$

\subsection{Analytical Theory by Houston Function}
\label{sec:3}

In this section we would like to show the derivation of the analytical formula employing the model Hamiltonian for the spatially periodic system,
\begin{equation}
\label{eq:TDSE-Otobe}
i\hbar\frac{\partial u_{n,\vec{k}}(\vec{r},t)}{\partial t}=\left(\frac{(\vec{p}+\hbar\vec{k}+\frac{e}{c}\vec{A}(t))^2}{2m}+V(\vec{r})\right)u_{n,\vec{k}}(\vec{r},t) .
\end{equation}
The time-dependent wave function $u_{n,\vec{k}}(\vec{r},t)$ can be expressed by the
Houston function\cite{Houston},
\begin{equation}
w_{n,\vec{k}}(\vec{r},t)=u_{n,\vec{k}+\frac{e}{c}\vec{A}(t)}^G(\vec{r})\exp\left[-\frac{i}{\hbar}\int^t dt' \epsilon_{n,\vec{k}+\frac{e}{c}\vec{A}(t')}^G \right],
\end{equation}
as $u_{n,\vec{k}}(\vec{r},t)=\sum_i C_i(t)w_{i,\vec{k}}(\vec{r},t)$.
Here, $\epsilon$ is the eigenenergy of the electron, $\vec{k}$ is the Bloch wavevector, $n$ in the band index, and $G$ indicates the ground state.

To simplify the system, we assume the parabolic two-band system defined as
\begin{equation}
\epsilon_{c,\vec{k}}-\epsilon_{v,\vec{k}}=\epsilon_{g}+\frac{\hbar^2|\vec{k}|^2}{2\mu}, 
\end{equation}
where $\epsilon_{g}$ is the band gap,  $\mu$ is the reduced mass, and $v(c)$ presents valence (conduction) band.
The Houston function can be expands by the $e^{-il\Omega t}$
\begin{equation}
\label{Houston}
w_{v(c),\vec{k}}(\vec{r},t)=\sum_l W^l_{v(c),\vec{k}}(\vec{r},t)e^{-il\Omega t},
\end{equation}
with continuous wave$\vec{A}(t)=\vec{A_0}\cos\Omega t$ \cite{otobe16,otobe16-2}.
Here, $W^l_{v(c),\vec{k}}(\vec{r},t)$ is the $l-$th order coeffient.
Since Eq.~(\ref{Houston}) corresponds to the expansion in to the dressed states (Floquet states) \cite{Mizumoto06},
the transient absorption can be understood as the response of the dressed states at time $T_p$.

The electronic current ($\vec{J}(t)$) is important to consider the optical response.
From the Fourier transformation of the current, 
we can estimate the conductivity as,  $\sigma(\omega)=\frac{\tilde{J}(\omega)}{\tilde{E}(\omega)}$,
where $\tilde{J}(\omega)$ is the Fourier component of current, and $\tilde{E}(\omega)$ is the applied field.
From Eq.~(\ref{eq:TDSE-Otobe}), the current is expressed as, 
\begin{eqnarray}
\vec{J}(t)&=&-\frac{n_ee^2}{mc}\vec{A}(t)-\frac{e}{2m}\sum_{n,\vec{k}}\left[\frac{1}{\Omega_{cell}}\int_{\Omega_{cell} } d\vec{r} u^*_{n,\vec{k}}(\vec{r},t)\vec{p}u_{n,\vec{k}}(\vec{r},t)+C.C.\right],
\end{eqnarray}
where $n_e$ is the electron density, $\Omega_{cell}$ is the volume of the unit cell.

We can derive the transient dielectric function with the usual linear response treatment under the elliptically polarized light,
$\vec{A}(t)=A_0(\eta\sin\Omega t,0,\cos\Omega t)~(0\le \eta \le 1)$, as

\begin{eqnarray}
\label{E_DFKE}
&&\varepsilon_E(T_p,\omega)=1-\frac{4\pi e^2}{m\omega^2}n_e-\frac{2 e^2|p_{cv}|^2\mu^{3/2}}{\sqrt{2}m^2\pi} \nonumber\\
&\times&\int_0^{\infty} \sqrt{\epsilon_k}d\epsilon_k \int _{-1}^{1} d\cos\theta \int_0^{2\pi} d\phi  \sum_{l_1,l_2,\zeta_1,\zeta_2} \nonumber\\
&\times&\frac{(-1)^{\zeta_1-\zeta_2}e^{i2\zeta_2\Omega T}}{\omega+2\zeta_2\Omega}\tilde{J}_{l_1}(\alpha,\beta)\tilde{J}_{l_1-2\zeta_1}(\alpha,\beta)\nonumber\\
&\times&\left[ \frac{J_{l_2}(\gamma)J_{l_2+2(\zeta_1-\zeta_2)}(\gamma)}{\omega-(\epsilon_g+\epsilon_k+U_E +(l_1+l_2-2\zeta_2)\Omega)} \right.\nonumber\\
&-&\left. \frac{J_{l_2}(\gamma)J_{l_2+2(\zeta_1+\zeta_2)}(\gamma)}{\omega+(\epsilon_g+\epsilon_k+U_E +(l_1+l_2+2\zeta_2)\Omega)}\right]
\end{eqnarray}
\cite{otobe16-2}.
Here, $\alpha, \beta, \gamma$ are defined as
\begin{eqnarray}
\alpha&=&\frac{e k A_0}{c\mu \Omega}\cos\theta,\\
\beta&=&\frac{e^2 A_0^2}{8c^2\mu \Omega}(1-\eta^2),\\
\gamma&=&\eta \frac{e k A_0}{c\mu \Omega}\sin\theta\cos\phi,
\end{eqnarray}
respectively,  $U_E$ is the ponderomotive energy, $J_l$ is the $l$-th order Bessel function, 
$\tilde{J}_l(a,b)$ is the generalized Bessel function \cite{Reiss03}, 
\begin{equation}
\tilde{J}_l(a,b)=\sum_{m}J_{l-2m}(a)J_{m}(b),
\end{equation}
and $p_{cv}$ is the transition moment between valence and conduction band.
The $\theta$ ($\phi$) is the angle between $\vec{k}$ and $z-$ ($x-$) axis, 
$\vec{k}=k(\sin\theta\cos\phi, \sin\theta\sin\phi,\cos\theta)$.

Transient spectroscopy using ultrashort pulses observes not only the absorption by the dressed states
, but also the phase difference between them.
Therefore, the oscillation of $\varepsilon_E(T_p,\omega)$ in probe time $T_p$ is derived from the energy difference between the dressed states.
The oscillation period of $\varepsilon_E(T_p,\omega)$ is an even multiple of the frequency of the pump light due to the symmetry of the system.

\subsection{Time-Dependent Density Functional Theory}

A theory describing the ground state of a multi-electron system is density functional theory (DFT).
Electronic states are obtained by solving the Kohn-Sham equation, which is the basic equation of DFT\cite{KS}.
On the other hand,  
the electron dynamics under the laser fields can be described by the time-dependent Kohn-Sham (TDKS) equation
\begin{equation}
\label{TDKS}
i\hbar\frac{\partial}{\partial t}u_{n,\vec{k}}(t)=\hat{H}(t)u_{n,\vec{k}}(t)
\end{equation}
\begin{equation}
\hat{H}(t)= \frac{1}{2m}\left(\vec{p}+\hbar\vec{k}+\frac{e}{c}\vec{A}(t)\right)^2+V_{ion}+V_H+V_{xc},
\end{equation}
based on the time-dependent density functional theory (TDDFT)\cite{Runge84}.
Here, $\vec{A}(t)$ is the vector potential, $V_{ion}$ is the coulomb potential from ions, $V_H$ is the Hartree potential, 
and $V_{xc}$ is the exchange-correlation potential.
TDDFT is a good approach to describe the nonlinear attosecond electron dynamics \cite{Krausz13, Sommer16, Schultze14, Lucchini16,George14}.

We approximate the time-evolution of the wave function by the 4-th order Taylor expansion time-evolution operator \cite{Yabana96,Bertsch00}.
The wave functions and the Hamiltonian are discretized with three-dimensional grid \cite{Chelikowsky94}.
The core electrons are neglected by employing the norm-conserve pseudopotential \cite{TM91,Kleinman82}.

\subsection{Tr-DFKE in Diamond}

We present the numerical results for a diamond by the TDDFT.
We assume a cubic unit cell including eight carbon atoms.
We descretize the unit cell with $24^3$ grid points, and the K-space with $22^3$ grid points.
We assume the adiabatic approximation, and employed a local density approximation for the exchange-correlation potential \cite{PZ81}.
The pump laser pulse is defined as the quasi-continuous wave, 
\begin{equation}
\vec{E}(t) = \hat{e} E_0 F_P(t) \sin \Omega t,
\end{equation}
with frequency of $\Omega=0.4$ eV.
The $F_P(t)$ is the envelope function to apply the laser field adiabatically.
The probe laser is the gaussian type pulse,
 \begin{equation}
\vec{E}_{p}(t) = \hat{e}_{p} f_{p}  \sin(\omega_{p} t)\exp\left(-\frac{(t-T_p)^2}{\zeta^2}\right),
\end{equation}
where $T_p$ is the probe time.
We assume the pulse duration $\zeta=0.7$ fs, peak field intensity $f_{p}=2.7\times10^{-3}$ MV/cm, and frequency of $\omega_p=5.5$ eV.
The intensity of probe field is weak so as to linear response is dominant. The frequency $\omega_p$ corresponds to the optical band gap  in out calculation.

\begin{figure}
\centering
\includegraphics[width=80mm]{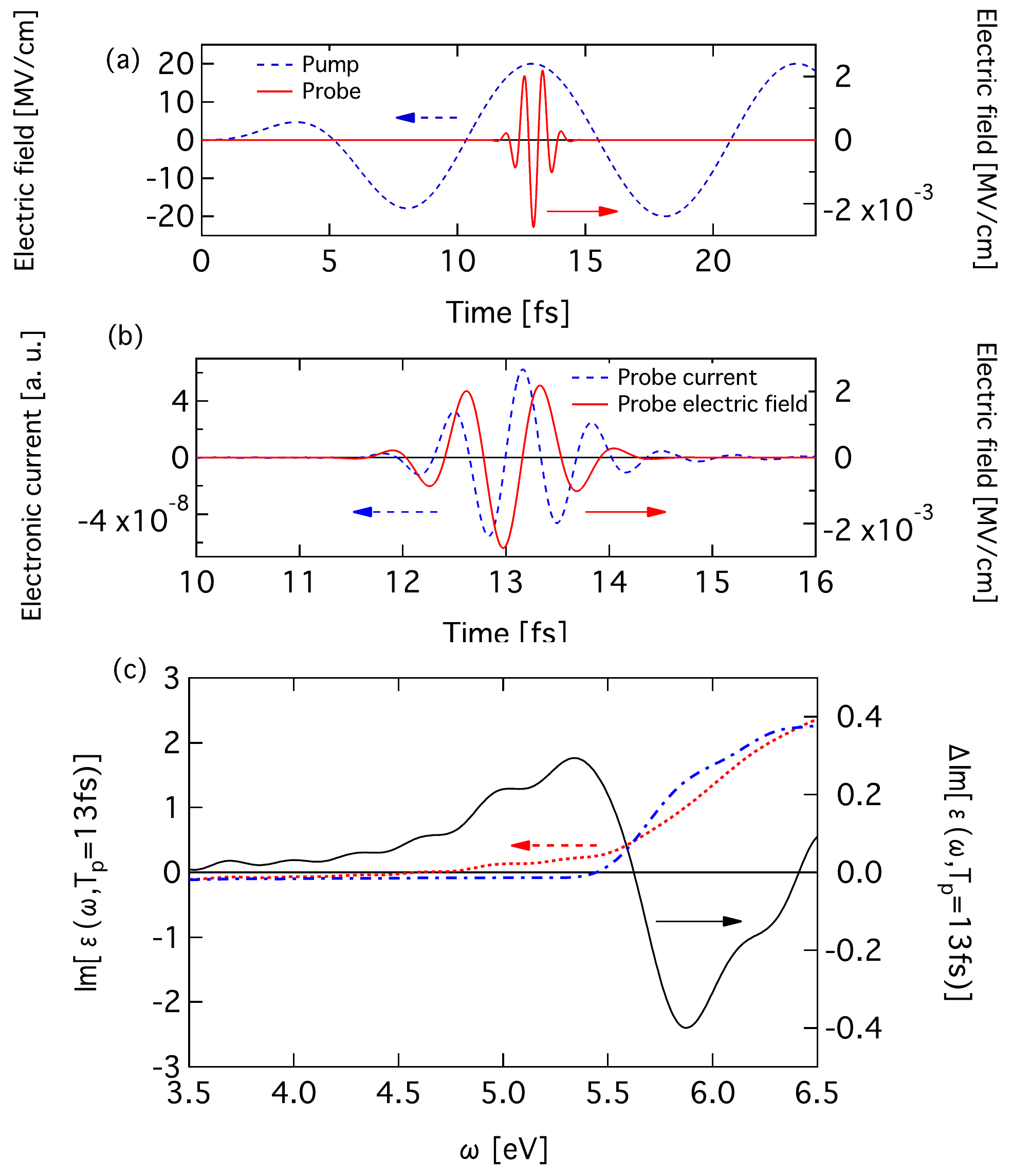} 
\caption{\label{fig1} (a) Pump (dashed line) and  
probe (solid line) electric fields are shown. 
(b) The electronic current (solid line)  
induced by the probe electric field (dashed line). 
(c) The imaginary part of the dielectric function in the 
presence of the pump field,  
${\rm Im}[\varepsilon(T_p=13fs,\omega)]$ (dashed line), 
and in the absence of the pump field,  
${\rm Im}[\varepsilon(\omega)]$ (dot line).  
Solid line shows the difference, 
${\rm Im}[\varepsilon(T_p=13fs,\omega)]- 
{\rm Im}[\varepsilon(\omega)]$.
Reprinted with permission from Ref.~\cite{otobe16}. Copyright 2016 by American Physical Society. 
 } 
\end{figure}

We apply the electric fields of $E(t$) (dashed line) and $E_p (t)$ (solid line) as shown in Fig. \ref{fig1} (a).
The field intensity of the pump light is 20 MV / cm, and the time of the probe $T_p$ is 13~fs.
The polarization of the pump (probe) field is parallel to [1,0,0] ([0.0.1]).
A solid line in Fig.~\ref{fig1}(b) indicates an electronic current density $J_p(t)$ induced by the $E_p(t)$.
We use atomic unit (a.u.) in all calculation.  
A dashed line in Fig.~\ref{fig1}(c) indicates the imaginary part of the dielectric function ${\rm Im}[\varepsilon]$, which is calculated from the Fourier component.
the $J_p(t)$ and $E_p(t)$.
For comparison, ${\rm Im}[\varepsilon]$ without  the pump 
is indicated by a dotted line, and the difference between them, 
$\Delta\rm{Im}[\epsilon(T_p,\omega)]$, is indicated by a solid line.
We can see the absorption below, and transparency above the band gap.

\begin{figure}
\centering
\includegraphics[width=90mm]{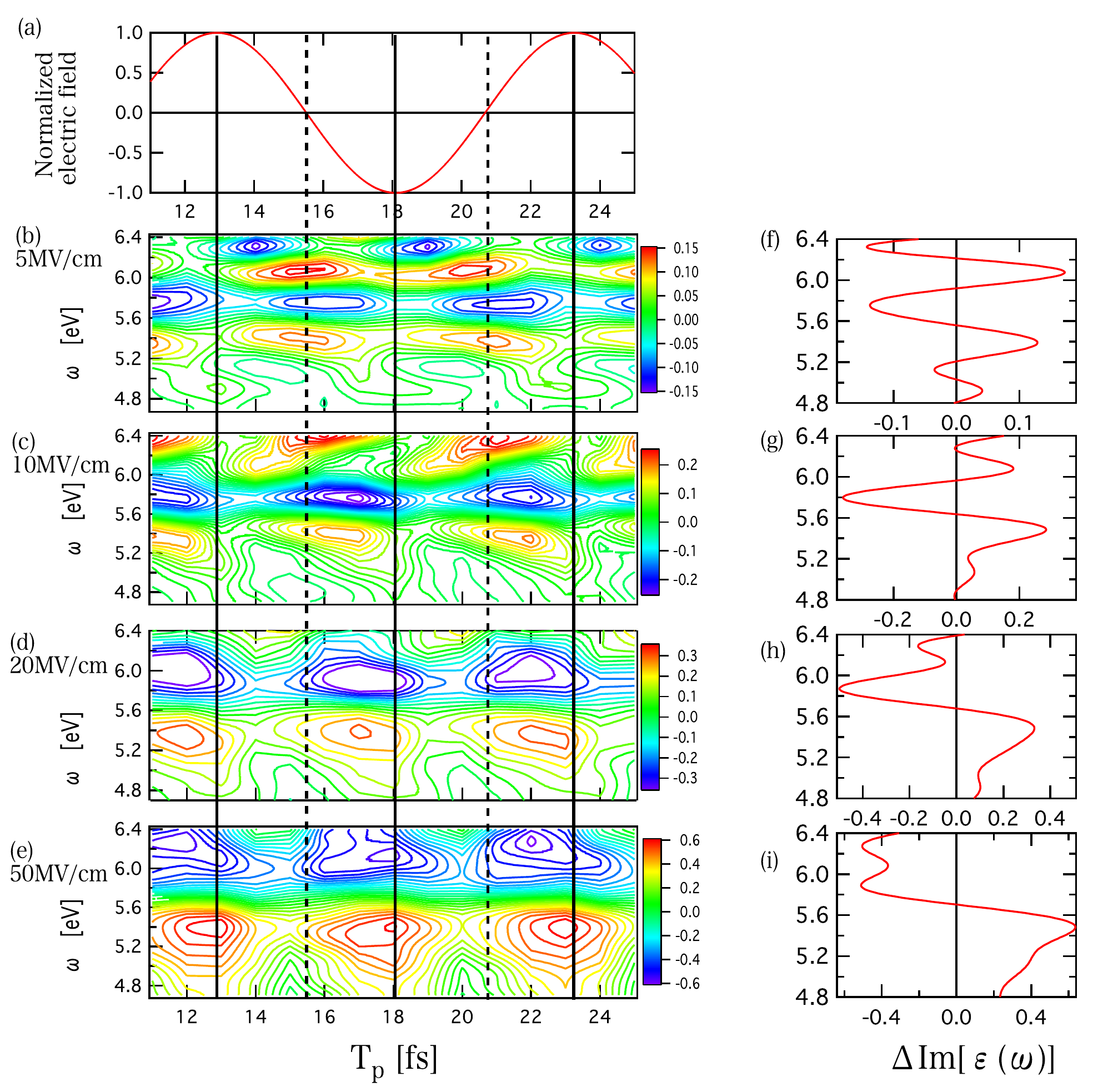} 
\caption{\label{fig2} Contour plots of $\Delta{\rm Im}[\varepsilon(T_p,\omega)]$ under  
the  pump field of the intensity of (b) 5, (c) 10, (d) 20, and (e) 50 MV/cm. 
(f)-(i) present the time average of the $\Delta{\rm Im}[\varepsilon(T_p,\omega)]$. 
The normalized pump field is  
shown in (a). The vertical solid lines indicate the time when the pump field is maximum, 
and the dashed lines indicate the minimum.  
Reprinted with permission from Ref.~\cite{otobe16}. Copyright 2016 by American Physical Society.}
\end{figure}

Fig. ~\ref{fig2} shows the probe time dependence of $\Delta\rm{Im}[\varepsilon(T_p,\omega)]$ and its dependence on peak intensity of the pump field
 as a function of probe time $T_p$ (fs) and frequency $\omega$.
In the case of $E_0=5$ MV/cm (Fig.~\ref{fig2}(b)), the  $\Delta\rm{Im}[\varepsilon(T_p,\omega)]$ show the maximum at the time 
when the pump light field is zero around the optical band gap.
The peak of the $\Delta\rm{Im}[\varepsilon(T_p,\omega)]$ at each $\omega$ shifts backward as the $\omega$ decreases from the band gap.
The peak of the $\Delta\rm{Im}[\varepsilon(T_p,\omega)]$ shifts backward as the pump field intensity increases.
In the case of $E_0=50$ MV/cm (Fig.~\ref{fig2}(e)), whereas the peak coincides with the peak of the pump field intensity, 
the $\omega$ dependent shift becomes weak.
The coincidence between the pump field and the $\Delta\rm{Im}[\varepsilon(T_p,\omega)]$ in Fig.~\ref{fig2}(e) indicates the adiabatic
response of the electrons.

\begin{figure} 
\centering
\includegraphics[width=90mm]{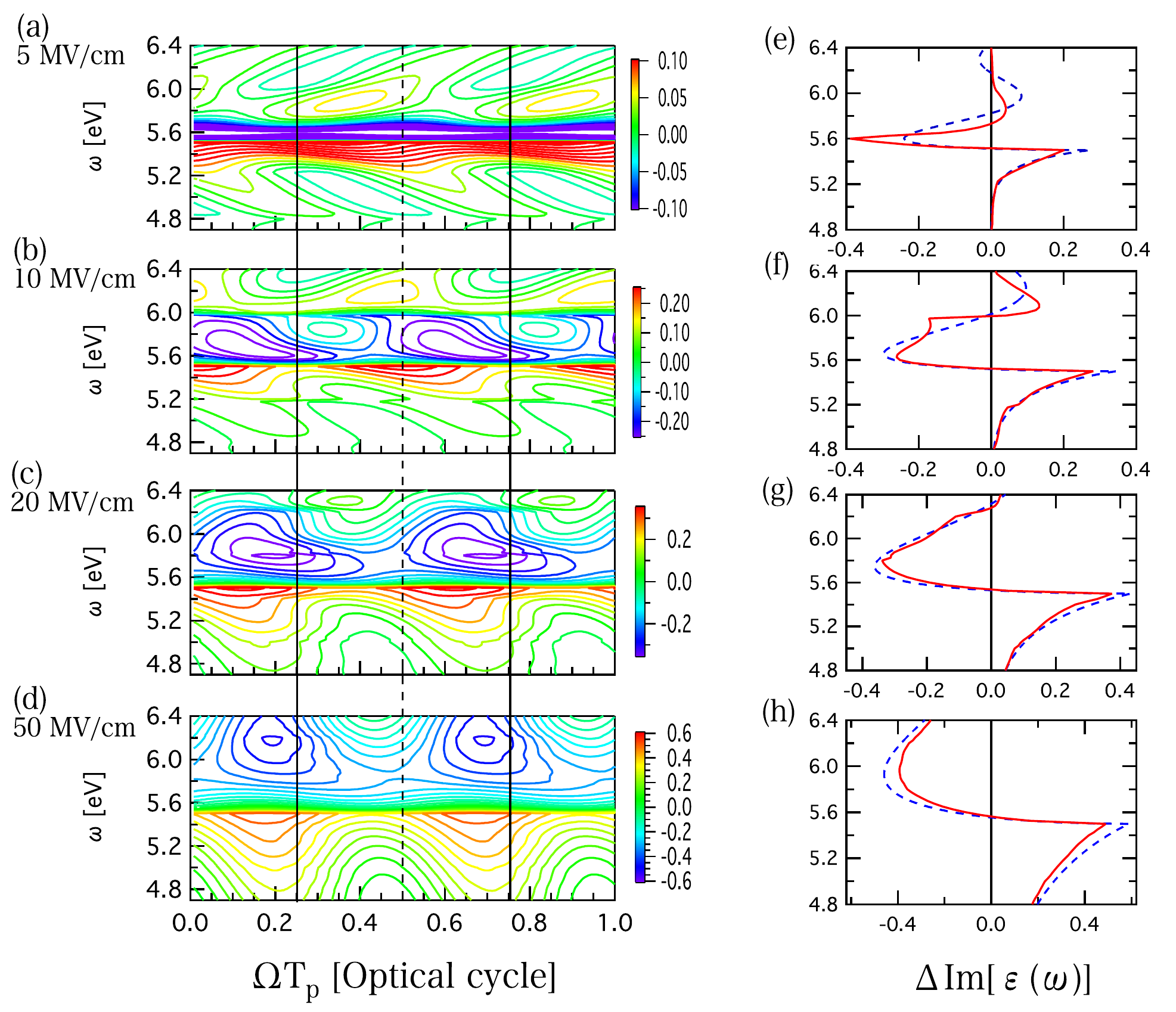} 
\caption{\label{fig3} Contour plots of 
$\Delta {\rm Im}[\varepsilon(T_p,\omega)]$ in the two-band model 
for the pump field intensities of (a) $E_0=5$, (b) 10,  (c) 20, and (d) 50 MV/cm.   
The horizon axis is the phase defined by $\Omega T_p$. 
(e)-(h) indicates the time average of the $\Delta {\rm Im}[\varepsilon(T_p,\omega)]$ (red solid lines) 
and the modulation assuming usual FKE.
Reprinted with permission from Ref.~\cite{otobe16}. Copyright 2016 by American Physical Society.}
\end{figure}

Let us now compare the TDDFT and analytical model calculations.
Fig.~\ref{fig3} shows the results by the model calculation \cite{otobe16,otobe16-2}.
We present the time depencen of $\Delta{\rm Im}[\varepsilon(T_p,\omega)]$ in left panels, and time-averaged modulation in right panels.
Whereas the peak of $\Delta{\rm Im}[\varepsilon(T_p,\omega)]$ coincident with the minimum of the pump field intensity,
 it shifts to the maximum of the pump field intensity as the field intensity increases.
The $\omega$ dependence is also qualitatively agree with Fig~\ref{fig2}.
The agreement of model calculation with the TDDFT results indicates that the Tr-DFKE can be understood by the 
response of the dressed states.
We also present in Fig.~\ref{fig3} (e)-(h) the modulation by the static electric field assuming the FKE.
The interesting point is that the DFKE and FKE show similar behavior as the pump laser field increases.

\begin{figure}
\centering
\includegraphics[width=80mm]{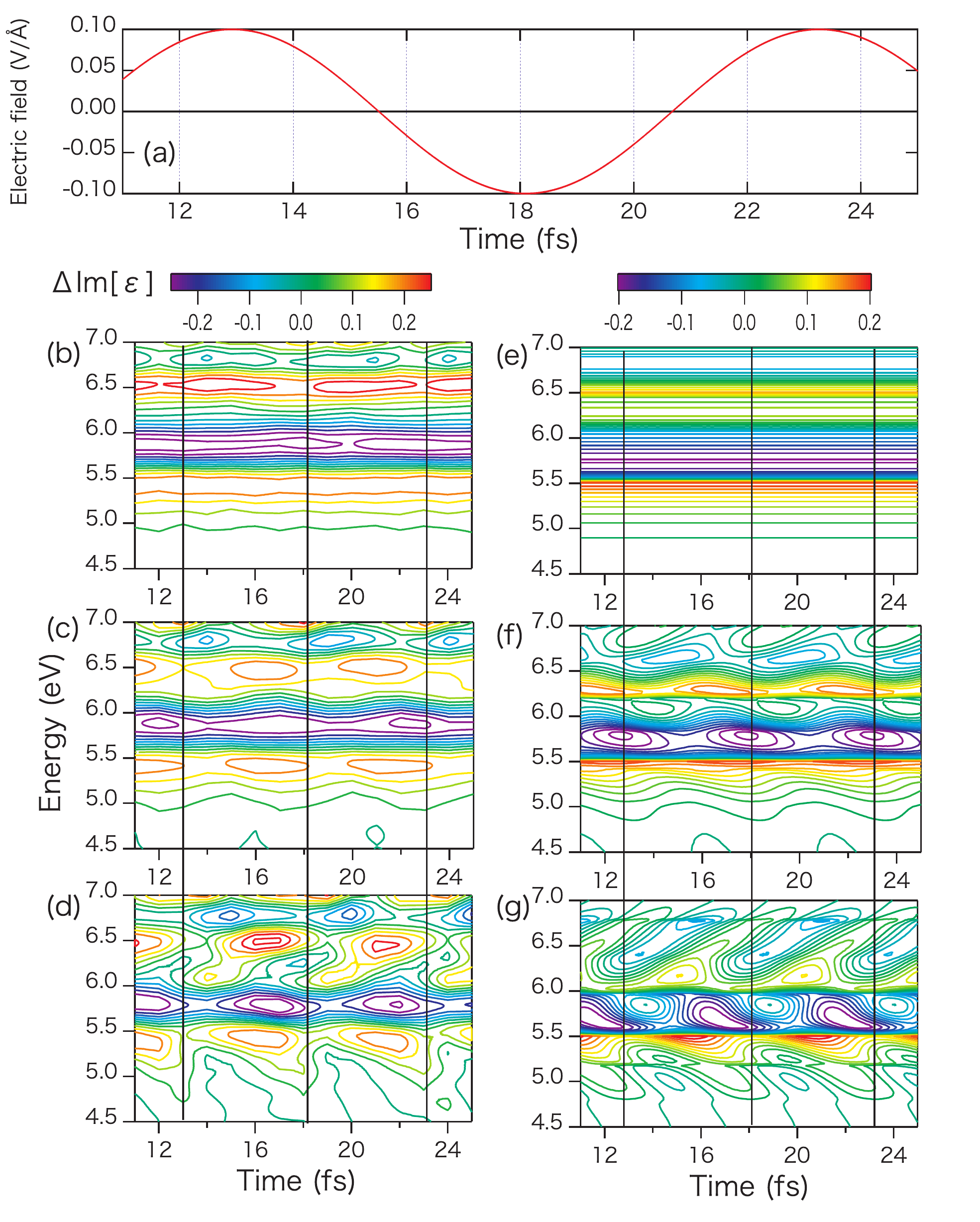}
\caption{\label{fig4}   $\Delta{\rm Im}[\varepsilon(T_p,\omega)]$  under circularly ((b) and (e)),  elliptically ((c) and (f)), and  linearly ((d) and (g)) polarized lasers.
(b)-(d) present the numerical results from TDDFT, and (e)-(g) are the results from 
 Eqs.~(\ref{E_DFKE}). 
  Panel (a) plots the electric field of the pump laser against the probe time, the range of which corresponds to that for 
the real-time TDDFT simulation. The vertical solid lines indicates the maximum of the pump field intensity.
Reprinted with permission from Ref.~\cite{otobe16-2}. Copyright 2016 by American Physical Society.}
\end{figure}

We would like to see the ellipticity ($\eta$) dependence in the next step.
Fig.~\ref{fig4} shows the $\eta$ dependence in TDDFT (left panels) and model calculation (right panels).
Fig.~\ref{fig4} (b) and (e) are the circularly polarization, (c) and (f) are elliptic polarization with $\eta=\sqrt{2}$, and (d) and (g) are linear polarization case.
We present only the positive modulation of $ {\rm Im}[\varepsilon(T_p,\omega)]$ in Fig.~\ref{fig4}.
As the reference, the pump light field with linear polarization is shown in Fig.~\ref{fig4}(a).
The maximum field intensity is set to 10 MV/cm for all calculations. 
The time-dependence of $\Delta{\rm Im}[\varepsilon(T_p,\omega)]$ becomes weak as the ellipticity increases in both case.
It should be note that the time-dependence disappears in circularly polarization.
In our model, we assume parabolic bands which is isotropic system.
The diamond also relatively isotropic system. 
Therefore, in the circularly polarization, the electron cannot distinguish the oscillation of the field in average. 

%
%
%

\section{Summary}
\label{sec:Summary}

We have compiled recent development in theoretical and numerical modeling of strong-field electron dynamics in solids.
First, we have the concept of introduced intraband and interband transitions, forming the basis for discussion in the momentum space, through the rigorous derivation for graphene.
Then, we have extended it to the multiband, momentum-space three-step model. The electron-hole interaction effects can also be incorporated in this model.
Moreover, we have presented the TD-DM and TDDFT methods for actual three-dimensional materials, whose predictions can be quantitatively compared with experimental results.

However, much more theoretical and experimental investigations are yet to be done, in order to reach comprehensive understanding of the electron dynamics in various solid-state materials subject to intense laser pulses.
It is expected that further accumulation of knowledge from different perspectives, such as suitable use and eventual unification of real-space and momentum-space pictures, effects of electron correlation, impurity, relaxation, and decoherence, and high-field phenomena in topological insulators and quantum materials, will lead to discovery of novel phenomena in various functional materials as well as innovative applications of high-intensity lasers.

\begin{acknowledgement}
This research was supported in part by a Grant-in-Aid for Scientific Research (Grants No. 23104708, No. 26390076, No. 17K05070, No. 18H03891, No. 18K14145, No. 19H02623, and No. 19H00869) from the Ministry of Education, Culture, Sports, Science and Technology (MEXT) of Japan and also by the Photon Frontier Network Program of MEXT. 
This research was also partially supported by the Center of Innovation Program (Grant No. JPMJCE1313) from the Japan Science and Technology Agency (JST), JST CREST (Grant No. JPMJCR15N1), MEXT Quantum Leap Flagship Program (Grant No. JPMXS0118067246), the Research and Education Consortium for Innovation of Advanced Integrated Science by JST, and the Exploratory Challenge on Post-K Computer from MEXT. 
The computation in this work was done using the facilities of the Supercomputer Center, the Institute for Solid State Physics, The University of Tokyo, and also using the K computer provided by the RIKEN Advanced Institute for Computational Science through the HPCI System Research project (Project ID: hp160260, hp170235, and hp180174).
\end{acknowledgement}

\end{document}